\documentclass[letterpaper, 10 pt, conference]{ieeeconf}

\IEEEoverridecommandlockouts               
\renewcommand{\baselinestretch}{0.98}

\usepackage[T1]{fontenc}
\usepackage{amssymb,amsmath,epsfig}
\usepackage[latin5]{inputenc}
\usepackage{graphicx,times,latexsym}
\usepackage{subfigure}
\usepackage{psfrag,cite}
\usepackage{setspace}
\usepackage{algorithmic, algorithm}
\usepackage{color}
\usepackage{url}
\newtheorem{theorem}{Theorem}
\newtheorem{lemma}{Lemma}
\newtheorem{remark}{Remark}
\newtheorem{assume}{Assumption}
\newtheorem{defn}{Definition}

\newtheorem{corollary}{Corollary}

\newcommand{\oprocendsymbol}{\hbox{$\bullet$}}
\newcommand{\oprocend}{\relax\ifmmode\else\unskip\hfill\fi\oprocendsymbol}

\newcommand{\blue}[1]{\color{blue}{#1}}

\definecolor{dark-green}{rgb}{0,.4,0}

\newcommand{\modulo}{\text{mod}}

\usepackage{mathtools}
\mathtoolsset{showonlyrefs=true}

\newcommand{\real}{\mathbb{R}}
\newcommand{\integers}{\mathbb{N}}

\newcommand{\floor}[1]{\lfloor #1 \rfloor}
\newcommand{\ceil}[1]{\lceil #1 \rceil}

\providecommand{\figref}{Fig.\,\ref}
\providecommand{\secref}{Sec.\,\ref}
\providecommand{\thmref}{Thm.\,\ref}
\providecommand{\lemref}{Lem.\,\ref}

\providecommand{\asumref}{Assumption\,\ref}

\allowdisplaybreaks

\graphicspath{{epsfiles/}}

\title{Theory and implementation of event-triggered stabilization\\ over digital channels}

\author{Mohammad Javad Khojasteh$^{\dagger}$, Mojtaba Hedayatpour$^{\dagger}$, Massimo Franceschetti
  \thanks{$^{\dagger}$ Equal contribution}
\thanks{M. J. Khojasteh and M. Franceschetti are with the Department of Electrical and Computer Engineering of University of California, San Diego. M. Hedayatpour is with DOT Technology Corp. Regina, Canada. 
(e-mails: \texttt{\{mkhojasteh,massimo\}@ucsd.edu, mojtaba@seedotrun.com})
}
}
 
\begin{document}
\maketitle
\begin{abstract}
In the context of event-triggered control, the timing of the triggering events  carries information about the state  of the system that can be used for stabilization. 
%In the presence of event triggered transmissions, the timing of the triggering events carries implicit information. 
At each triggering event, not only can information be transmitted by the message content (data payload) but also  by its timing. 
%It is possible to exploit this timing information along with the message content (data payload)
%F
%Transmission of information, not only by message content (data payload) but also with its timing, is an example of these and is the topic of this paper. 
%In the same way that subsequent pauses in spoken language are used to convey information, it is also possible to transmit information in communication systems not only by message content (data payload), but also with its timing. 
We demonstrate this in the context of stabilization of a laboratory-scale inverted pendulum around its 
%unstable 
equilibrium point over a digital communication channel with bounded unknown delay.
%fact 
Our event-triggering control strategy   encodes timing information by transmitting in a state-dependent fashion and can achieve stabilization  using a data payload transmission rate  smaller than what the data-rate theorem prescribes for classical periodic control policies that do not exploit timing information.
Through experimental results, we show that as the delay in the communication channel increases, a higher data payload transmission rate is required to fulfill the proposed event-triggering policy requirements.
%In addition, our results also 
This confirms  the theoretical intuition that a larger delay brings a larger uncertainty about the value of the state  at the controller, as less timing information is carried in the communication. Our results also provide a novel encoding-decoding scheme to achieve input-to-state practically stability (ISpS) for nonlinear continuous-time systems under appropriate assumptions.
\end{abstract}
%\begin{keywords}
 %  Control under
 % communication constraints, event-triggered control, %networked control systems,
 % quantized
 % control %topological feedback entropy.
%\end{keywords}

\section{Introduction}\label{sec:intro}
Event-triggered control has gained significant attention due to its advantages    over conventional control schemes  in cyber-physical systems (CPS). 
%The recent technological advances in wireless communications and computation, and their integration into cyber-physical systems (CPS), open the door to a myriad of new, challenging, and exciting problems and opportunities to engineers. 
Although periodic control is the most common and perhaps simplest solution for digital systems, it can be inefficient in sharing communication and computation resources~\cite{Tabuada,wang2011event,WPMHH-KHJ-PT:12,pearson2017control,tallapragada2013event,postoyan2015framework,girard2015dynamic,yildiz2019event,linsenmayer2018containability,seuret2016lq}. 
%astrom2002comparison%tallapragada2018event
%Due to its simplicity periodic control has been prevalent in digital control systems. However, in the context of CPS~\cite{kumar}, periodic control does not efficiently utilize the distributed  communication and computation resources. This led to the advent of event-triggered control~\cite{astrom2002comparison,Tabuada,WPMHH-KHJ-PT:12}. 
The central concept of event-triggered control is to transmit sensory data only when needed to satisfy the control objective. 
%The main idea behind event-triggered control is to include value of the state for the determination of the times at which data-transmissions take place in a control loop. 
In addition to utilizing the distributed resources efficiently, it has been proven that carefully crafted event-triggered policies outperform the linear-quadratic (LQ) performance of the periodic control policies~\cite{khashooei2018consistent}.
%antunes2017consistent
Another advantage of event-triggered control  is that the timing of the triggering events, effectively revealing the state of the system, carries information that can be used for stabilization.  This allows achieving stabilization with a transmission rate lower than that required by periodic control strategies  \cite{khojasteh2017time,khojasteh2018stabilizing,OurJournal1}.
 
In networked control systems a finite-rate digital communication channel closes the loop between the sensor and the controller. In this setting, data-rate theorems~\cite{Nair,Mitter,khina2019control,kostina2016rate11,Massimo,martins2006feedback,hespanha2002towards} provide the communication channel requirements for stabilization. 
\begin{figure}[t]
	\centering
 \includegraphics[scale=0.35]{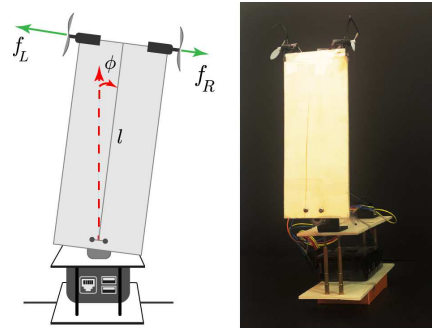}
	\caption{An inverted pendulum controlled by thrust force of two propellers. The pendulum is a plywood sheet of length $l$. The angle $\phi$ of the pendulum from the vertical line and its rate of change, measured by the sensor and transmitted to the controller over a digital channel with bounded unknown delay, are used  to determine
    %is controlled 
    the left and right thrust forces $f_L$ and $f_R$ of the propellers. }\label{fig:robot} 
\end{figure}
%In communication theory, there are different approaches towards information transmission in digital channels. A rather new method of transmitting information is adjusting the transmission time of a symbol compared to conventional methods of adjusting amplitude of a signal~\cite{khojasteh2018stabilizing,anantharam1996bits,rose2016inscribed1}. 
%That being said, a finite-rate digital communication channel can be described as two parallel channels used for data payload and timing information transmission. 
%Needless to say, data-rate theorems have been evolved around these approaches. These  theorems generally state that 
They state that to ensure stabilization of an unstable linear system,  the minimum information rate communicated over the channel, including both data payload and timing information, must be at least  equal to the \emph{entropy rate} of the plant, defined as the sum of the unstable modes in nats~\cite{khojasteh2017time,khojasteh2018stabilizing}. 
When information is encoded in the timing of the transmission events   using event-triggered control, our previous work~\cite{princeton_paper,khojasteh2018event21122}  has shown the existence of an event-triggering strategy that achieves input-to-state practically stability (ISpS)~\cite{jiang1994small,sharon2012input}  for
%can
%practically 
%stabilize % (resulting in the states to be bounded for all times beyond any fixed horizon)
%(requiring the state to be bounded at all times beyond a fixed horizon)
any linear, time-invariant system  subject to bounded disturbance over a digital communication channel with bounded delay using a data payload transmission rate lower than the entropy rate. 
%(In the case that the initial condition, delay, and system disturbances are bounded the stability notion used in~\cite{princeton_paper,khojasteh2018event21122} requires the state to be bounded at all times
%beyond a fixed horizon. It is easy to verify the results in~\cite{princeton_paper,khojasteh2018event21122} are also valid for the general notion of ISpS.)
This is possible because,
for small values of the delay, the timing information is substantial, and the data payload transmission rate can be lower than the entropy rate of the plant. However, as the delay increases, a higher data payload transmission rate is required to satisfy the requirements of the proposed event-triggering control strategy.

A similar data-rate theorem formulation also holds for nonlinear systems.
%The majority of results on control under communication constraints are restricted to linear plants, and 
The  works~\cite{Topological,liberzon2005stabilization,de2005n} for
nonlinear systems are restricted to plants without disturbances and
with a bit-pipe communication channel.  The work~\cite{Topological} uses the entropy of topological dynamical systems to elegantly
determine necessary and sufficient bit rates for local uniform asymptotic stability. Consequently, the results are only local and
derived under restrictive assumptions. Under appropriate assumptions, the
work~\cite{liberzon2005stabilization}  extends to nonlinear but
locally Lipschitz systems, the zoom-in/zoom-out strategy
of~\cite{liberzon2003stabilization}. The sufficient
condition proposed in this work is, however, conservative, and
does not match the necessary condition proposed in~\cite{Topological}. The work~\cite{sharon2012input} further extend the results in~\cite{liberzon2005stabilization} to linear systems with uncertainty and under appropriate assumptions to nonlinear systems with disturbances.
Inspired by the Jordan block decomposition employed in~\cite{Mitter}
to design an encoder/decoder pair of a vector system, the
work~\cite{de2005n} provides a sufficient design for feed-forward
dynamics that matches the necessary condition proposed
in~\cite{Topological}. The recent work
in~\cite{sanjaroon2018estimation} studies the estimation of a
nonlinear system over noisy communication channels, providing a
necessary condition over memoryless communication channels and a
sufficient condition in case of additive white Gaussian noise channel.
%The recent work
%in~\cite{sanjaroon2018estimation} studies the estimation of a
%nonlinear system over noisy channels, providing a
%necessary condition over memoryless communication channels and a
%sufficient condition in case of additive white Gaussian noise channel.
%over an erasure 
%Our contributions are twofold. 

The majority of results on control under communication constraints are restricted to theoretical works. 
%Most of the literature in the area of control over   communication  channels in the past two decades has been restricted to theoretical results. 
Here for the first time, we examine   data-rate theorems in a practical setting, using an inverted pendulum, a classic example of an inherently unstable nonlinear plant with numerous practical applications. Our first contribution is to implement the event-triggering control design introduced in~\cite{princeton_paper,khojasteh2018event21122}, and demonstrate the utilization of timing information to stabilize a laboratory-scale inverted pendulum   over a digital communication channel with bounded unknown delay, see~\figref{fig:robot}. 
%~\cite{anderson1989learning}. 
%kafetzis2017inverted %kajita20013d %grasser2002joe
A video that illustrates the main ideas and
demonstrates our experimental results 
%for the laboratory-scale inverted pendulum in \figref{fig:robot} 
can be found at
{\blue{
\url{https://youtu.be/1P0i-tWsPoA}}}. 
The results of our experiments show  that using the sufficient packet size derived in~\cite{princeton_paper,khojasteh2018event21122} on a linearized  model of the inverted pendulum around its unstable equilibrium point, the  state estimation error is sufficiently small and we can stabilize the system. We show that for small values of the delay the experimental data payload transmission rate is lower than the entropy rate of the plant. On the other hand, by increasing the upper bound on the delay in the communication channel,   higher data payload transmission rates are required to satisfy the requirements of the proposed control strategy. 
The  event-triggering policy developed in~\cite{princeton_paper,khojasteh2018event21122} can only stabilize the pendulum \emph{locally} around its equilibrium point, where linearization is possible. Our second contribution is to address nonlinear systems directly, and develop a novel event-triggering scheme that exploits timing information to render  a class of continuous-time nonlinear systems subject to disturbances ISpS.
%Also, for small values of the delay the experimental data payload transmission rate is smaller than entropy rate of the plant. On the other hand, by increasing the upper bound on the delay in the communication channel, higher data payload transmission rates are required to satisfy the requirements of the proposed control strategy. %The intuition at the basis of this result is that  a larger delay introduces more uncertainty in the state estimation process, as it provides access to less timing information. Results in this paper validate the theory presented in our previous works, and demonstrate the value of the timing information in a practical setting, showing the possibility of stabilization at a  data payload transmission rates substantially lower than what predicted by the data-rate theorem in a time-triggered scenario.
%Our experiments show that as delay in communication channel get larger, higher data payload transmission rate is required to fulfill the presented event-triggering policy requirements. This is true because with larger delay corresponds to  more uncertainties about the value of state at the controller increases, and less informative timing timing. 

From the system's perspective, our set-up is closest to the one in~\cite{liberzon2005stabilization,sharon2012input}, as  we consider locally Lipschitz nonlinear systems that can
be made input-to-state stable (ISS)~\cite{sontag2008input} 
with respect to the state estimation error and system disturbances. Using our encoding-decoding scheme, we encode the information in timing via event-triggering control in a  state-dependent fashion to achieve input-to-state practical stability (ISpS) in the presence of unknown but bounded delay. We also discuss the different approaches to eliminate the  ISS assumption. %While as~\cite{liberzon2005stabilization,sharon2012input} in general our design might be conservative, but for sufficiently small delay, we can beat the necessary condition proposed by data-rate theorem~\cite{de2005n} using the implicit timing information in event-triggering.

Finally, we point out that the work~\cite{dou2018sufficient} studies  event-triggering stabilization of globally Lipschitz nonlinear system without disturbances where the communication delay is arbitrarily small. Also, the  work~\cite{tanwani2017stabilization} investigates  event-triggered stabilization of nonlinear system under communication constraints but it does not explicitly quantify the effect of quantization in the presence of system disturbances, nor the timing information carried by the  triggering events. In addition, the recent work~\cite{music2018design} utilizes a time-triggered controller to stabilize a two-wheeled inverted pendulum around its upright position over  IEEE 802.11g (WiFi). Since the whole bandwidth of this channel is devoted to a two-wheeled inverted pendulum, the work~\cite{music2018design} does not explicitly examine the effect
quantization.
 
A complete list of notations and  proofs of all the results appear in the Appendix.

\section{System model}\label{sec:model}
%This section presents the mathematical modeling of the networked control system, the plant (inverted pendulum controlled by two propellers), and event-triggered control design for stabilizing the system. A more detailed description of the proposed event-triggered control strategy, encoding-decoding scheme and mathematical proofs can be found in our previous work~\cite{princeton_paper,khojasteh2018event21122}. 

\begin{figure}[h]
	\centering
 \includegraphics[scale=0.35]{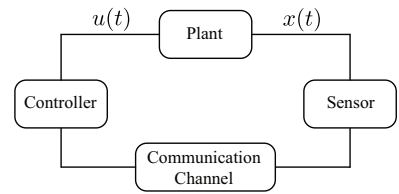}
	\caption{System model.}\label{fig:system} 
\end{figure}

\begin{figure}[t]
	\centering
 \includegraphics[scale=0.35]{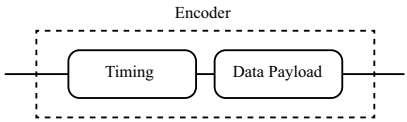}
	\caption{Representation of information transmission using data payload and transmission time of the packet in a digital channel. The encoding process consists of choosing the data payloads and their transmission times. Here, the sensor determines the transmission time using our event-triggering strategy in a state-dependent manner.
    %Using our event-triggering policy in addition to data
    %payload, the reception time of the packets transmitted over
    %the channel also carry information. Consequently, the communication channel can be seen as two parallel 		channels that transmit information in timing and data payload.
}\label{fig:timing channel} 
\end{figure}

We consider the  stabilization of the inverted pendulum depicted in \figref{fig:robot} around its unstable equilibrium point. The sensory information for stabilization is sent to the controller over a digital channel. The block diagram of the control system is given in \figref{fig:system}.
We assume the communication channel is capable of transmitting   packets composed of a finite number of bits   without error. Each transmitted packet is subject an unknown delay  upper bounded by $\gamma \geq 0$. 
In addition to the  data payload, the transmission time of the packets sent over the channel could be utilized to convey information to the controller.
As a result, the encoding process consists of choosing the timing and data payload of the packet, as shown in \figref{fig:timing channel}. 
%the timing information and data payload of the packet, as shown in Figure~\ref{fig:timing channel}.
In other words, in the sensor block, the quantized version of the state is encoded in a packet 
containing data payload as well as its timing.
%as well as its timing timing information.
In our design, the sensor encodes information in timing via an event-triggering technique in a state-dependent fashion.
%and is prepared for transmission to the controller 
%The explicit way to encode information in timing via an event-triggering technique and the packet construction is described in section~\ref{sec:controller}. 

At each triggering event, occurring at times $\{ t_s^k \}_{k \in \integers}$, the sensor transmits a packet $p(t_s^k)$ of length $g(t_s^k)$ over the communication channel. 
Packets arrive at the controller at times $\{ t_c^k \}_{k \in \integers}$. When referring to a generic triggering or reception time, we skip the super-script $k$ in $t_s^k$ and $t_c^k$.
%The decoding stage also happens after arrival of the packets in the controller block. 

Since the delay in the communication channel is upper bounded by $\gamma \geq 0$, the \emph{communication delays} represented by $\Delta_k = t_c^k-t_s^k$ ($k \in \integers$) must satisfy 
\begin{align}\label{gammma}
 \Delta_k\leq \gamma.
\end{align}
%If we define the size of the packet transmitted from sensor to the controller at each triggering time $t_s^k$ as $g(t_s^k)$ bits, 
By defining the $k^{th}$ \emph{triggering interval} as $\Delta'_k = t_s^{k+1}-t_s^{k}$, the \textit{information transmission rate} (the rate at which sensor transmits data payload over the channel) can be defined as 
%data payload over the channel is defined
\begin{align}\label{ratemo232oe4e}
  R_s = \limsup_{N \rightarrow \infty} \frac{\sum_{k=1}^{N} g(t_s^k)}{\sum_{k=1}^{N}
    \Delta'_k}.
\end{align}
%\magenta{
%The frequency with which transmission events are triggered is captured by the \emph{triggering rate}
%\begin{align} \label{trate}
%R_{tr} &= \limsup_{N\rightarrow
%    \infty}\frac{N}{\sum_{k=1}^N \Delta'_k}.
%%\end{align}
%}

%We denote by $\{ t_s^k \}_{k \in \integers}$ the sequence of  times when the sensor transmits a (data payload) packet of length $g(t_s^k)$ bits that contains a quantized version of the state. We let the   $k^{th}$ \emph{triggering interval} as $\Delta'_k = t_s^{k+1}-t_s^{k}$. The packets are delivered to the controller without error and entirely but with unknown upper bounded delay. Let $\{ t_c^k \}_{k \in \integers}$ be the sequence of times where the controller receives the packets transmitted at times $\{ t_s^k \}_{k \in \integers}$. We assume that the \emph{communication delays} $\Delta_k = t_c^k-t_s^k $, for all $k \in \integers$, satisfy 
% \begin{align}\label{gammma}
%  \Delta_k\leq \gamma ,
% \end{align}
% where $\gamma$ is a non-negative real number. At each triggering time, $t_s^k$ the sensor sends $g(t_s^k)$ bits, hence the \textit{information transmission rate } which is the the are at which sensor transmit 
% data payload over the channel is defined
% \begin{align}\label{ratemo232oe4e}
%   R_s = \limsup_{N \rightarrow \infty} \frac{\sum_{k=1}^{N} g(t_s^k)}{\sum_{k=1}^{N}
%     \Delta'_k}.
% \end{align}

\subsection{Plant Dynamics}
We consider a linearized version of the two-dimensional problem of balancing an inverted pendulum with two propellers, where the motion of the pendulum is constrained in a plane and its position can be measured by an angle $\phi$ representing small deviations from the upright position of the pendulum, as depicted in \figref{fig:robot}. 
The inverted pendulum has mass $m_1$ and length $l$. The propellers are identical and are attached to two motors of mass $m_2$. 
$m$ and $I$ respectively represent the total mass of the system and its moment of inertia.   Therefore, a nonlinear equation of the system can be written as follows
\begin{equation}
\label{nonlinear_sys}
I\ddot{\phi} = mgl\sin\phi(t)+ \xi(t)l + \textrm{noise},
%- b\dot{\phi} 
\end{equation}
where $g$ is the gravitational acceleration,
%, $b$ is the friction coefficient of the joint connecting the pendulum to its base
%For simplicity we assume the friction is small and can be neglected ($b\approx0$). 
and $\xi(t)$ is the resultant thrust force of the propellers ($f_L$ and $f_R$ as shown in \figref{fig:robot}) generating a moment about the axis of rotation of the pendulum. 
Note that in this nonlinear equation   the effect of the friction is included in the additive noise.
The force $\xi(t)$ can be estimated as a linear function of the control input $\tilde{u}(t)$, applied to the motors, with some proportionality constant $k_\xi$ (found from experiments), namely $\xi(t)=k_\xi\tilde{u}(t)$.

We derive the linearized equations of motion using a small angle approximation. This linearization is only valid for sufficiently small values of the delay upper bound $\gamma$ in the communication channel. Linearizing~\eqref{nonlinear_sys} around the equilibrium point results in the following dynamics
\begin{equation}
\label{linear_sys}
I\ddot{\phi} = mgl\phi(t) + k_{\xi}l\tilde{u}(t) + \textrm{noise}.
\end{equation}
By defining the state variable $\tilde{\pmb{x}}=(\phi, \dot{\phi})^T$,
the state-space equations can be written as follows
\begin{equation}
  \label{sys_dyn}
 \dot{\tilde{\pmb{x}}}=\tilde{\textbf{A}}
  \tilde{\pmb{x}}+ \tilde{\textbf{B}}\tilde{u}(t) + \tilde{\pmb{w}}(t),
  %\dot{\pmb{x}} 
  %= \textbf{A}\pmb{x} + \textbf{B}u + \pmb{w}(t)
\end{equation}
where
\begin{equation}
\textstyle
  \tilde{\textbf{A}} = 
  \begin{bmatrix}
  0 & 1 \\
  \frac{mgl}{I} & 0
  \end{bmatrix}, 
  \tilde{\textbf{B}} = 
  \begin{bmatrix}
  0 \\
  \frac{k_{\xi}l}{I}
  \end{bmatrix}.
\end{equation}
In our prototype  shown in \figref{fig:robot}, the pendulum is a plywood sheet of size $0.18\times0.073\times0.005$ m and mass $m_1=0.030$ kg. The motors are of mass $m_2=0.010$ kg. Also, $l=0.180$ m, and $g=9.81$ m/s$^2$. Using first principles, one can find the moment of inertia of the pendulum about its axis of rotation to be $I=3.57\times10^{-4}$ kg/m$^2$. 
By experiments, we approximate $k_\xi=0.001$.
 Therefore, the system~\eqref{sys_dyn} can be rewritten as follows
\begin{equation}
\textstyle
  \label{sys_dyn1}
  \dot{\tilde{\pmb{x}}} = 
  \begin{bmatrix}
  0 & 1 \\
  53.58 & 0
  \end{bmatrix}
  \tilde{\pmb{x}} + 
  \begin{bmatrix}
  0 \\
  0.50
  \end{bmatrix} \tilde{u}(t) + 
  \tilde{\pmb{w}}(t).
\end{equation}
Using~\eqref{sys_dyn} it follows $\tilde{w}_1(t)=0$. Also, by experiments we deduce $|w_2(t)|$ 
%for 
%i \in \{1,2\}$ 
is upper bounded by $0.02$.
%Using~\eqref{sys_dyn} it follows $\tilde{w}_1(t)=0$. Also, the second element of the noise vector  
%,  and the noise in the system 
%is characterized by experiments and is found 
%to be upper bounded as: $|\tilde{w}_2(t)|\le 0.02$.

Now using the eigenvector matrix %\textbf{P} 
\begin{equation}
\textstyle
	\label{P_matrix}
    \textbf{P} = 
    \begin{bmatrix}
    0.1354 & -0.1354 \\
    0.9908 & 0.9908
    \end{bmatrix}
\end{equation}
of matrix $\tilde{\textbf{A}}$ we consider a canonical transformation to diagonalize the system~\eqref{sys_dyn1} as follows
\begin{equation}\label{AAmj}
    \dot{\pmb{x}} = \textbf{A}\pmb{x}(t)+\textbf{B}u(t)+\pmb{w}(t),
\end{equation}
where $\textbf{A} = \textbf{P}^{-1}\tilde{\textbf{A}}\textbf{P}$, $\textbf{B}=\textbf{P}^{-1}\tilde{\textbf{B}}$, $\pmb{x}(t) = \textbf{P}^{-1}\tilde{\pmb{x}}(t)$
and $\pmb{w}(t) = \textbf{P}^{-1}\tilde{\pmb{w}}(t)$.  Therefore, for the diagonalized system~\eqref{AAmj} we have 
\begin{equation}
\textstyle
\label{Atilde}
\textbf{A} = 
\begin{bmatrix}
\lambda_1 & 0 \\
0 & \lambda_2
\end{bmatrix} = 
\begin{bmatrix}
7.3198 & 0 \\
0 & -7.3198
\end{bmatrix}, 
\end{equation}
\begin{equation}
\textstyle
\textbf{B} = 
\begin{bmatrix}
0.2523 \\
0.2523
\end{bmatrix},\,\,
\label{Btilde}
\pmb{x} = 
\begin{bmatrix}
3.6940\phi + 0.5046\dot{\phi} \\
0.5046\dot{\phi} - 3.6940\phi
\end{bmatrix}, 
\end{equation}
\begin{equation}
\textstyle
|w_i(t)| \le M=0.0470 \,\, \:\textrm{for}\: i \in \{1,2\},
\end{equation}
where the upper bound $M$ on the $|w_i(t)|$ for $i \in \{1,2\}$ is found by taking the maximum of upper bounds of all the elements in $\pmb{w}(t)$. 

% We consider, which is defined as follows.
 We now define a modified version of input-to-state practically stablity (ISpS)~\cite{jiang1994small,sharon2012input}, which is suitable for our event-triggering setup with the unknown but bounded delay in the digital communication channel.
\begin{defn}\label{cd12emkmkd}
The plant~\eqref{AAmj}
is 
ISpS if both of its coordinates $x_1(0)$ and $x_2(0)$ are ISpS. 
Also, $x_1(t)$ is ISpS if
there exist $\beta \in \mathcal{KL}$, $\psi \in \mathcal{K}_\infty(0)$, $d \in \real_{\ge 0}$,  $\chi \in \mathcal{K}_\infty(d)$, $d' \in \real_{\ge 0}$ and $\zeta \in \mathcal{K}_\infty^2(0,d')$  such that for all $ t\ge 0$
\begin{align}
    |x_1(t)| \! \le \! \beta\left(|x_1(0)|,t\right)\!+\!\psi\left(|w_1|_t\right)\!+\!\chi(\gamma)\!+\!\zeta(|w_1|_t,\gamma).
    %\sup_{s \in [0,t]} |w_1(s)|
\end{align}
\end{defn}
Note that, for a fixed $\gamma$, this definition reduces to the standard notion of ISpS. Given that the initial condition, delay, and system disturbances are bounded, ISpS implies that the state must be bounded at all times beyond a fixed horizon. 
%When $\gamma=0$, ISpS becomes input-to-state stability (ISS)~\cite{sontag2008input}. 
%In addition when the disturbance is zero ISS is equivalent to global asymptotic stability (GAS)~\cite{lin1996smooth}.  
%for any $x_1(0)$ in a closed interval,
%there exists an increasing function $\alpha_1$ of $M$, with $0\le\alpha_1(0)<\epsilon$ for any $\epsilon>0$, such that for all $\Psi_1>\alpha_1(M)$, there exists $T_1$ such that, $|x_1(t)|\le \Psi_1$ for all $t\ge T_1$.
%\end{defn}
%\begin{remark} {\rm 
%In the case that the initial condition, delay, and system disturbances are bounded, ISpS requires the state to be bounded at all times
%beyond a fixed horizon.}
%  \oprocend
%\end{remark}

Since $\lambda_2$ in~\eqref{AAmj} is negative, the second coordinate is inherently stable, and we do not need to transmit updates about the second coordinate to the controller via the communication channel. However, since $\lambda_1$ is positive, the uncertainty about the first coordinate grows exponentially at the controller, hence  the sensor needs to communicate information to the controller about the state of the first coordinate to render the plant ISpS~\cite{princeton_paper}. 
%Also, to achieve ISpS for system~\eqref{AAmj} and given the architecture of the system described in Figure~\ref{fig:system},
%encoding-decoding scheme, the quantized version of the states as a packet and the packet size $\{g(t^k_s)\}_{k \in \integers}$ to guarantee practical stability 
%To ensure the dynamics~\eqref{AAmj} is practically stable given the limitations posed by the system model described in Figure~\ref{fig:system} the sensor should, select the sequence of transmission times $\{ t_s^k \}_{k \in \integers}$ and the packet sizes $\{g(t^k_s)\}_{k \in \integers}$ which is the subject of the next section. 

A brief description of the event-triggered control approach  in our previous work~\cite{princeton_paper,khojasteh2018event21122} which determines the sequence of transmission times $\{ t_s^k \}_{k \in \integers}$ and  packets $\{p(t^k_s)\}_{k \in \integers}$ to achieve ISpS for the first coordinate of the dynamics~\eqref{AAmj} is available at Appendix~\ref{sec:controller}.

\section{Implementation and System Architecture}\label{sec:architect}
\begin{figure}[t]
  \centering
  \includegraphics[scale=0.52]{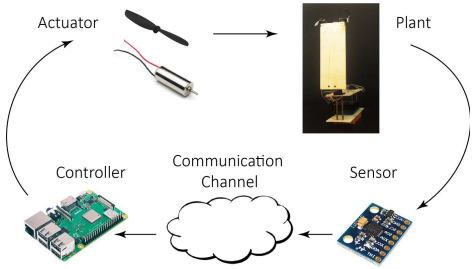} 
  \caption{Architecture and components of the prototype. }
  \label{fig:components}
\end{figure}

We now present the details of the implementation of the proposed event-triggered control scheme on a real system, along with experimental results validating the theory. 
The prototype used  is an inverted pendulum system built using off-the-shelf components. The body of the system is made of plywood sheets, as depicted in \figref{fig:robot}. For sensors, we use InvenSense MPU6050 MEMS sensor which has a 3-axis accelerometer and a 3-axis gyroscope, and we use a complementary  filter to read the angle and angular velocity of the pendulum. We choose Raspberry Pi Model 3 for the computation unit and the controller in the system. For actuation, we use two small DC motors equipped with two identical propellers. \figref{fig:components} depicts the different components of the system. 

Using the plant dynamics introduced in~\eqref{AAmj}, we implement the event-triggered control scheme proposed in Appendix~\ref{sec:controller} on the prototype system. 
While our theory is developed for continuous-time plants, the experiments are performed on digital systems and in discrete-time domain with small enough sampling time~$\delta$ to make the discrete-time model as close to the continuous-time model as possible. Because of this discretization, the minimum upper bound for the channel delay is equal to two sampling times. A delay of at most one sampling time exists from the time that a triggering occurs to the time that the sensor takes a sample from the plant state and another delay of at most one sampling time exists from the time that the packet is received to the time the control input is applied to the plant. 
In the experiments, a triggering occurs as soon as $z_1$ is equal or greater
than $J$ and the controller
 has received the previous packet, in this way since the sampling time is
small, at the triggering time, equation~\eqref{eq:ets} will be valid approximately.
%Also, The packet size for the experiments has two differences from the lower bound provided in~\eqref{packet_size}. In digital systems, the packet size in an integer greater than or equal to one, therefore, we use the ceiling operator in~\eqref{packet_size}, and we set the minimum size of the packet to one. 

To simulate the digital channel between the sensor and the controller, we send packets composed of a finite
number of bits from the sensor to the controller with a delay, that is a multiple of the sampling time $\delta$, upper bounded by $\gamma$. 
%These updates are encoded and quantized in a packet of size $g(t_s^k)$ found in~\eqref{packet_size}. 

\section{Experimental results}\label{sec:expsetup}
\begin{figure*}[t]
\centering
\begin{tabular}{c c c}
	 & \scriptsize{$M=0.0470$, $\gamma=2\delta=0.006$ sec, $g(t_s)=1$ bit} & \\
    \includegraphics[width=60mm]{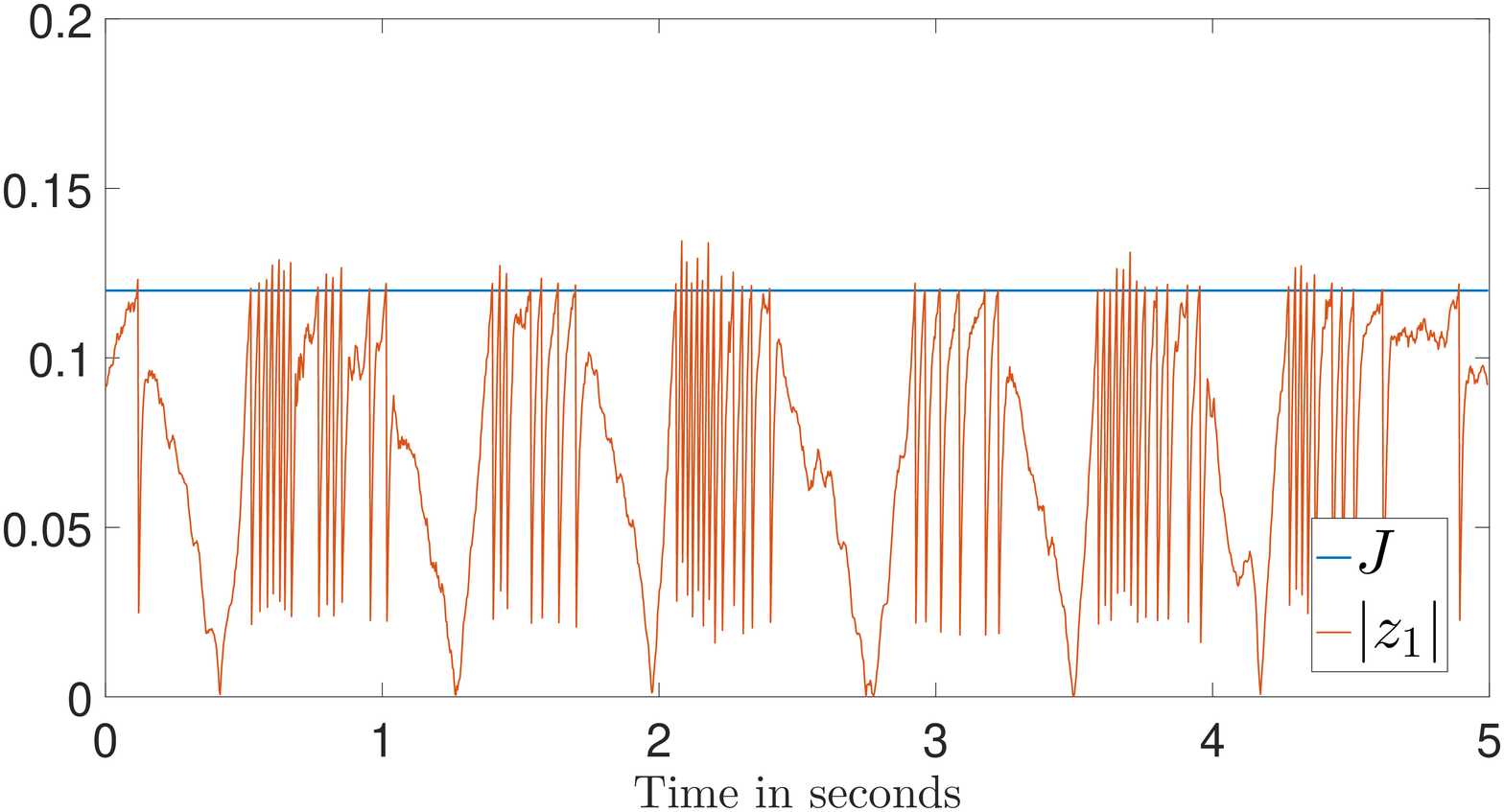} &
    \includegraphics[width=60mm]{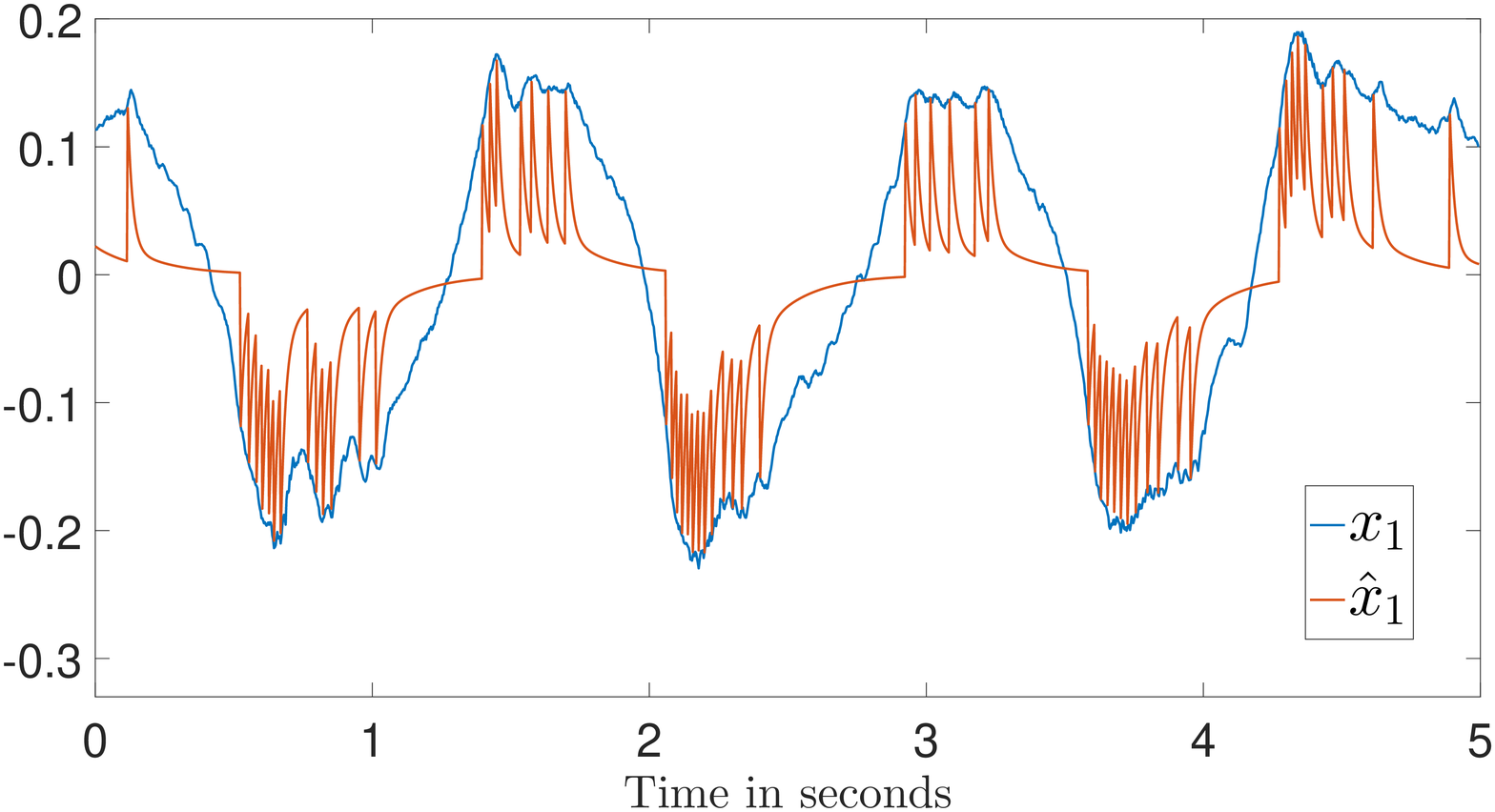} & 
    \includegraphics[width=60mm]{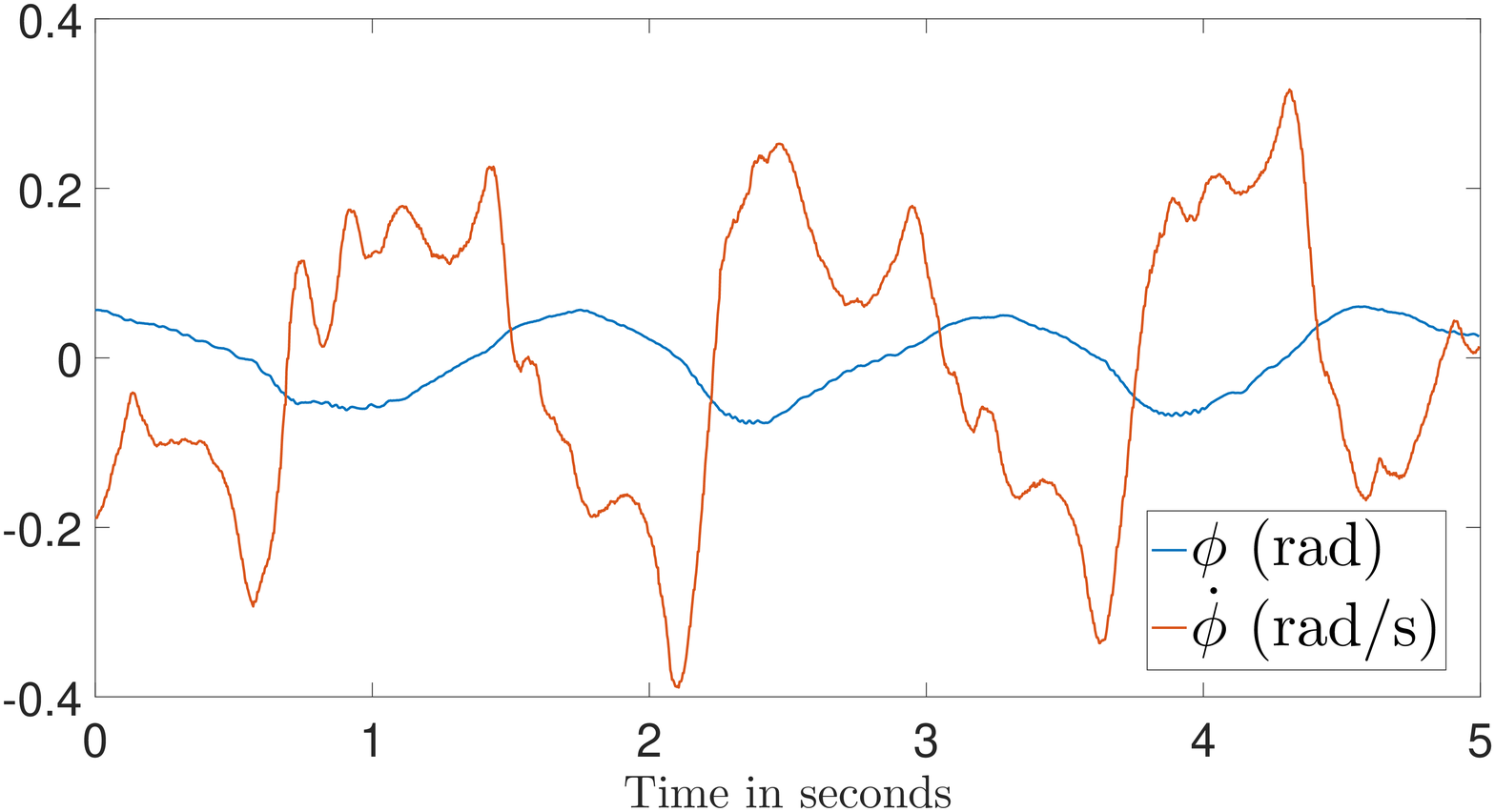} \\
     & \scriptsize{$M=0.0470$, $\gamma=5\delta=0.015$ sec, $g(t_s)=7$ bits} & \\
    \includegraphics[width=60mm]{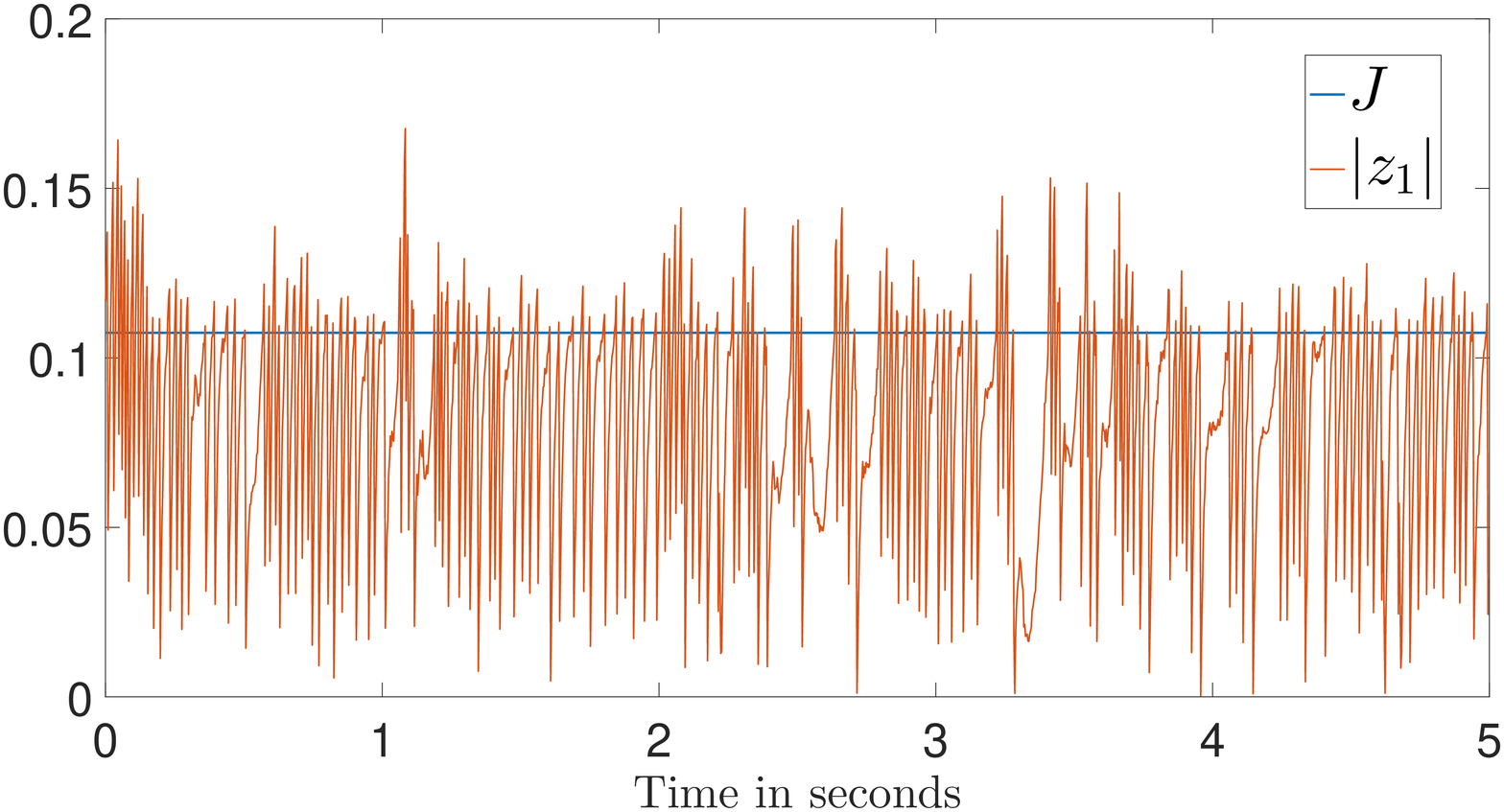} &
    \includegraphics[width=60mm]{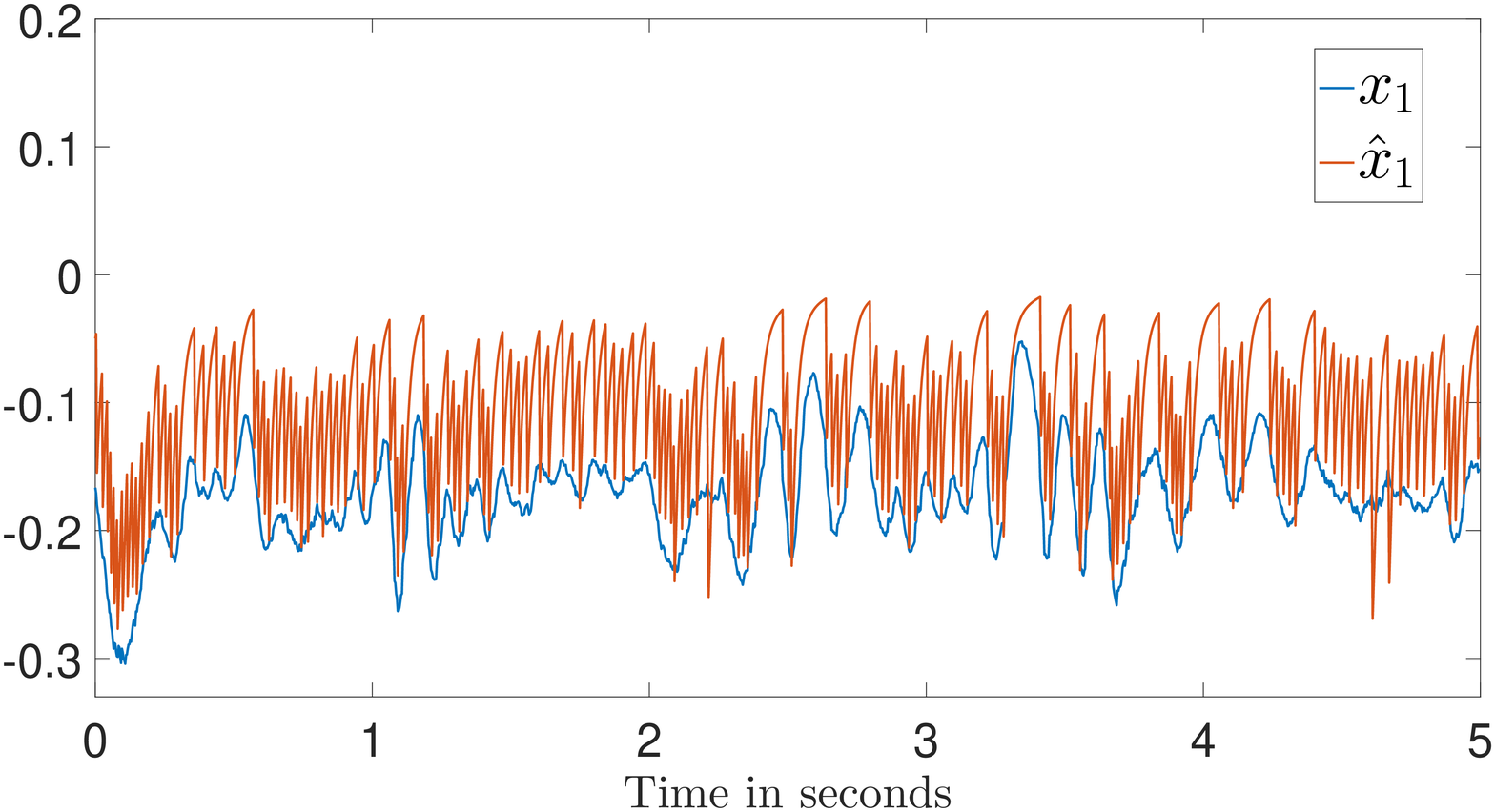} &
    \includegraphics[width=60mm]{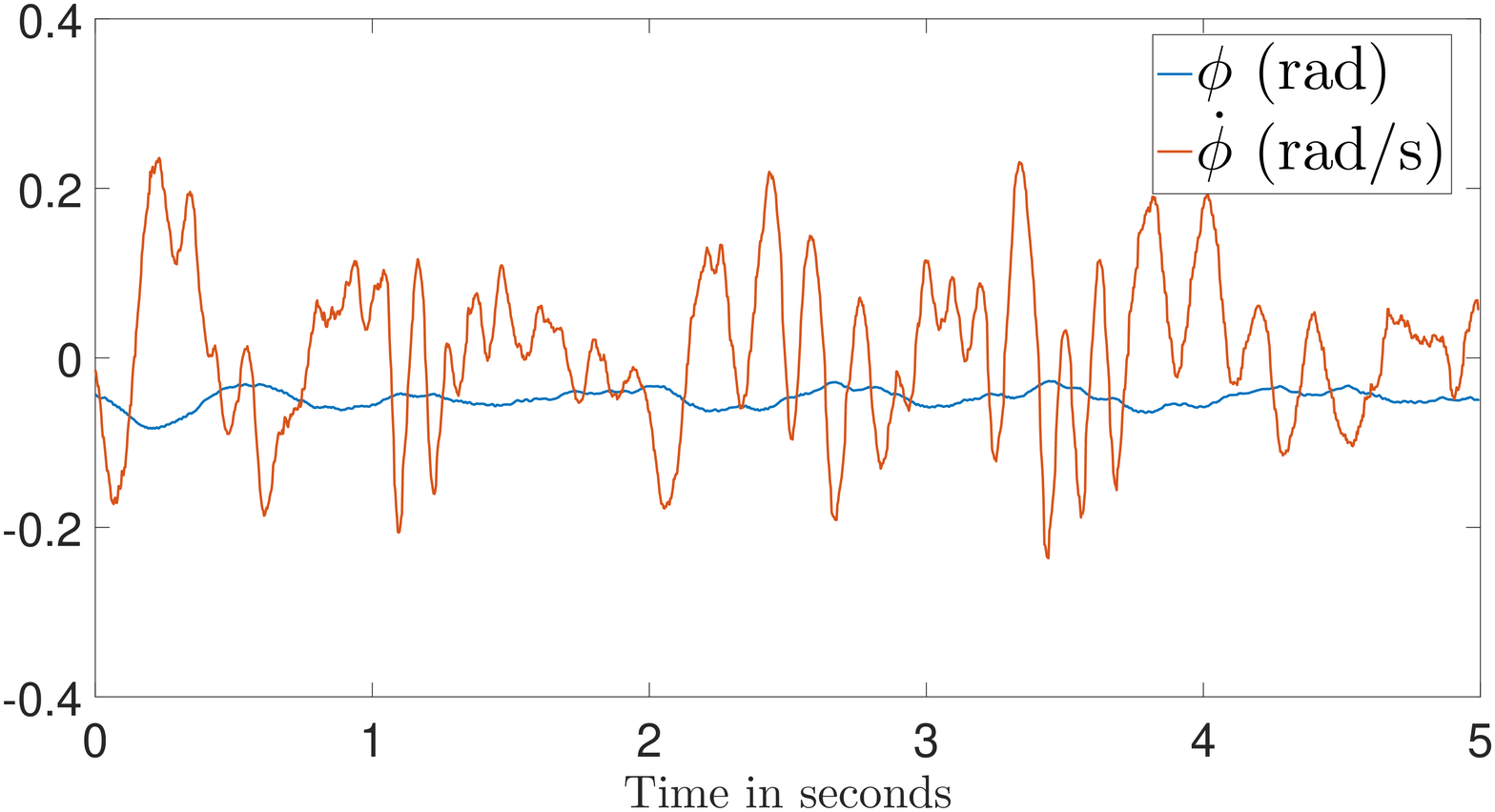} \\
\end{tabular}
\caption{Experimental results for stabilizing the inverted pendulum over a digital channel with random delay upper bounded by two sampling times (first row) and five sampling times (second row). When $\gamma=2\delta$, the packet size is 1 bit and when $\gamma=5\delta$, the packet size becomes 7 bits. 
}
\label{exp_res}
\end{figure*}

In this section, experimental results for various scenarios are presented. In all the experiments, the sampling time $\delta$ is 0.003 seconds, which is the smallest sampling time that the  measurements from our sensors permit. 
Also we set $\rho_0=0.01$, $b=1.00001$, and $J=\frac{M}{\lambda_1\rho_0}(e^{\lambda_1\gamma}-1)+0.1$. In the first set of experiments, we evaluate the performance of the controller for different values of $\gamma$. In \figref{exp_res}, the first row presents the results when $\gamma=0.006$ seconds or two sampling times and the second row presents the results when $\gamma=0.015$ seconds or five sampling times. The first column is the evolution of the absolute value of the state estimation error~\eqref{eq:state-estimation-error} (red) in time along with the triggering threshold (blue). As the absolute value of this error is greater than or equal to the triggering function, a triggering occurs and the sensor transmits a packet to the controller. However, due to the random delay (upper bounded by $\gamma$) in the communication channel, this error could grow beyond the triggering function. 
%However, because there is random delay in the channel, this error grows beyond the triggering function exponentially with this delay upper bounded by $\gamma$.
This growth, of course, can become larger as $\gamma$ increases which is shown in the first column of \figref{exp_res}. The first column also shows, more triggering occurred when the channel delay is upper bounded with five sampling times. 
%as the delay increases, the number of triggering events increases. 

The second column in \figref{exp_res} presents the evolution of the state $x_1$ (blue) corresponding to the unstable pole in the diagonalized system~\eqref{AAmj} and its estimate $\hat{x}_1$ (red) in time. The last column shows the evolution of the actual states of the system, namely the angle of the pendulum in radians and its rate of change in radians/sec. It can be seen that $|\phi|$ remains less than $0.2$ radians which ensures the linearization of~\eqref{nonlinear_sys} remains valid and is a good approximation. 

\begin{figure}[t]
  \centering
  \includegraphics[scale=0.16]{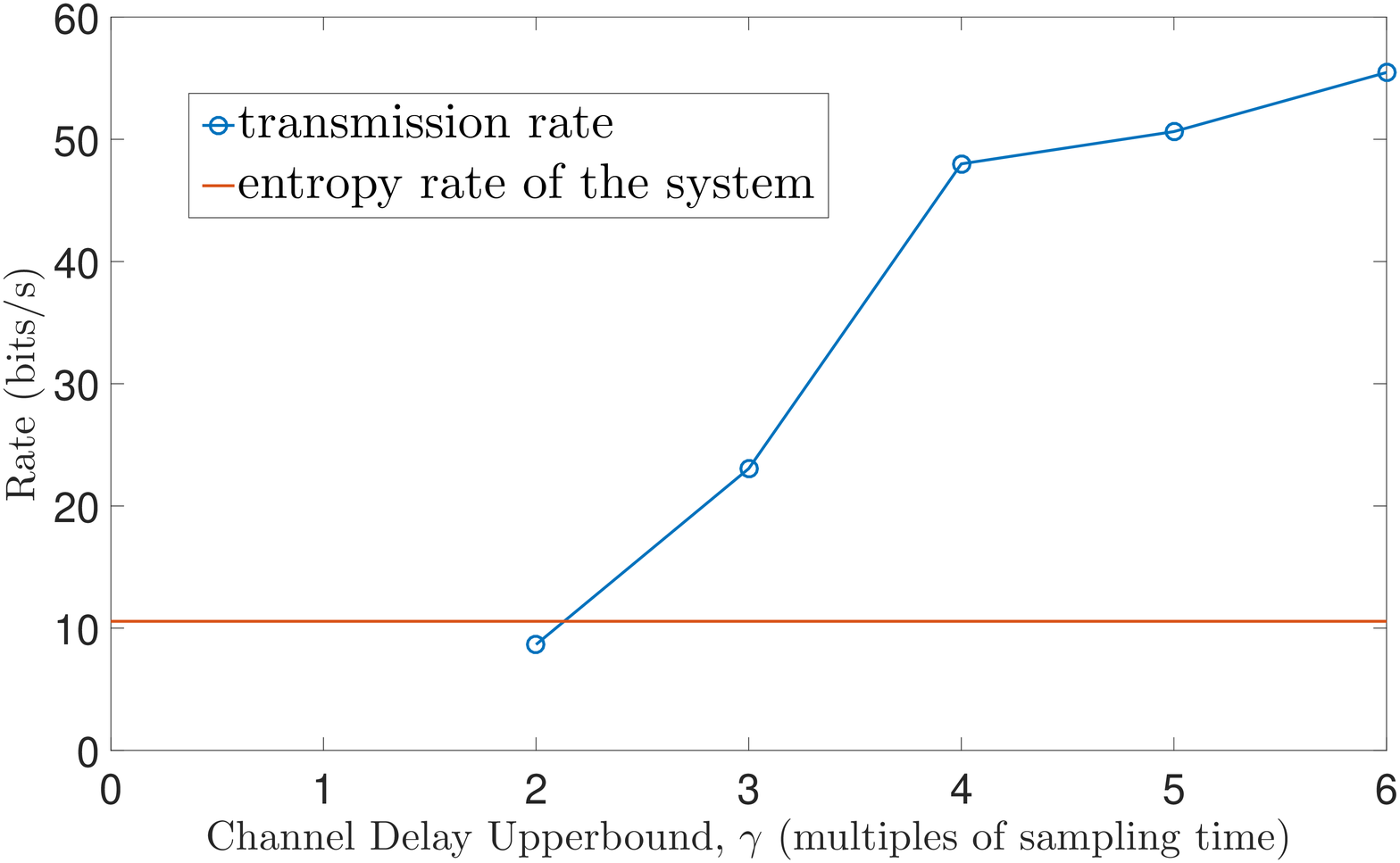} 
  \caption{Information transmission rate in experiments compared with the entropy rate of the system. Note that the rate calculated from experiments does not start at zero worst-case delay because the minimum channel delay upper bound is equal to two sampling times (0.006 seconds). The entropy rate of the system is equal to $\lambda_1/\ln2=10.56$ bits/sec while the minimum transmission rate for worst-case delay equal to two sampling time in the experiments is equal to 8.66 bits/sec.}
  \label{rate}
\end{figure}

We repeat the experiments for different values of $\gamma$ and calculate the sufficient transmission rate using~\eqref{ratemo232oe4e}.
According to  the data-rate theorem, to stabilize the plant, the information rate communicated over the channel in data payload and timing should be larger than the entropy rate of the plant~\cite{khojasteh2018stabilizing,khojasteh2017time}. In our experiments, when $\gamma=2\delta$ the timing information is substantial, therefore, the information transmission rate becomes smaller than the entropy rate of the plant which is shown in \figref{rate}. Furthermore, according to the theory developed in~\cite{princeton_paper,khojasteh2018event21122} as $\gamma$ increases, more information has to be sent via data payload for stabilization  since larger delay corresponds to more
uncertainties about the value of the states at the controller and
less timing information.
%According to theory, for small values of the delay we should be able to stabilize the system with any positive transmission rate and this rate can be less than entropy rate of the system as well which occurs . As $\gamma$ increases, more information has to be sent for stabilization which increases transmission rate to values higher than the entropy rate of the system and it can also be observed in the results shown in Figure~\ref{rate}. 

%In addition to  the data payload, the reception time of the packets carries information. Consequently,
%Likewise, 
%let $b_c(t)$ be the amount of information measured in bits included in data payload and timing information %number of bits that 
%received at the controller until time $t$.   
%The \emph{information access rate} is
%\begin{align*}
%  R_c = \limsup_{t \rightarrow \infty} \frac{b_c(t)}{t}.
%\end{align*}

%According to the data-rate theorem, if $R_c < A/\ln2$, the value of the state in~\eqref{syscon} becomes unbounded as $t\rightarrow \infty$  (the result for plants evolving in continuous time stated in~\cite[Theorem~1]{hespanha2002towards} does not consider disturbances, but can readily be generalized to account for them),
%and~\eqref{syscon} is not practically stable.
%In our discussion below, the data-rate theorem serves as a baseline for comparison with our results on the information transmission rate $R_s$ to better understand the amount of   timing information contained in event-triggered control designs in presence of unknown communication delays.

%Finally, the robustness of the controller is evaluated against additional disturbances and the results are shown in Figure~\ref{ext_disturbance}. The additional disturbances are applied to the system at time $t=2$ seconds and the evolution of $|z_1|$, $x_1$ and $\hat{x}_1$ in time are presented. It can be seen that even in presence of additional disturbances which are quite large, the event-triggered control policy is able to stabilize the system.
%
%\begin{figure}[b]
%  \centering
%  \includegraphics[scale=0.19]{sim3z.eps}
%  \includegraphics[scale=0.19]{sim3diag.eps}
%  \caption{Robustness of the event-triggered control strategy against additional disturbances. }
%  \label{ext_disturbance}
%\end{figure}
%
\setlength{\tabcolsep}{1pt}

\begin{remark} {\rm 
Similar to our analysis in~\cite{princeton_paper}, we assume the plant disturbance is random but bounded. In most of our experiments, we successfully stabilized inverted pendulum around its equilibrium point. Disturbances outside the prescribed limits occur rarely, but can still happen occasionally. Assuming that the disturbances are unbounded one might be able to extend the second-moment stability results of~\cite{nair2004stabilizability} to our setup.  %which is weaker than the practical stability that we studied here. 
Similarly, the case where the delay in the communication channel becomes unbounded with a positive probability is another interesting research problem.
  }
  \oprocend
\end{remark}

\section{Extension to nonlinear systems}\label{sec:nonlinear1}
The results developed in~\cite{princeton_paper,khojasteh2018event21122} are restricted to linear systems, and they can only stabilize the pendulum~\eqref{nonlinear_sys} locally, where the linear approximation is valid. Thus, now we develop a novel event-triggering scheme that encodes information in timing and under appropriate assumptions renders a continuous-time nonlinear system with disturbances ISpS. Clearly, the results of this section compare to the results of~\cite{princeton_paper,khojasteh2018event21122} are more sophisticated to analyze and implement.
%Thus, they are less sophisticated to develop and implement. However, 
%Consequently, in the previous sections, we linearized the nonlinear dynamics of the inverted pendulum around its equilibrium point. Therefore, the developed event-triggering policy 

 We consider sensor, communication
channel, controller system depicted in \figref{fig:system}, and  a continuous nonlinear plant 
\begin{align}
\label{nonlinearsystem1}
    \dot{x}=f(x(t),u(t),w(t)),
\end{align}
where $x$, $u$, and $w$ are real numbers representing the plant state, control input, and plant disturbance. Furthermore, we assume that for all time $t \ge 0$
\begin{align}
\label{disturbupperbounnonlinearcase}
    |w(t)| \le M.
\end{align}
%where the system disturabces 
%$w(t) \le M $
%The system in~\eqref{nonlinearsystem1}  is input-to-state practically stable (ISpS)~\cite{jiang1994small,sharon2012input}  if there exist $\beta \in \mathcal{KL}$, $\psi \in \mathcal{K}_\infty$, and $\chi \in \mathcal{K}_\infty$  such that for all $ t\ge 0$
%\begin{align}
%    |x(t)|\le \beta\left(|x(0)|,t\right)+\psi\left(\sup_{s \in [0,t]} |w(s)|\right)+\chi(\gamma).
%\end{align}
%When $\gamma=0$, ISpS becomes input-to-state stability (ISS)~\cite{sontag2008input}. 
%~\cite{sontag1989smooth}

As in~\eqref{sysest}, the controller constructs the state estimation $\hat{x}$, which  evolves during the inter-reception times as
\begin{align}
\label{nonlinearsystemestime}
    \dot{\hat{x}}=f(\hat{x}(t),u(t),0) \quad t \in (t_c^k,t_c^{k+1}),
\end{align}
starting at $\hat{x}(t_c^{k+})$  that is constructed by the controller using information received up to time $t_c^{k+}$. 
The explicit way to construct $\hat{x}(t_c^{k+})$ will be explained later in this section (see~\eqref{estimNonlin}).
As discussed in Appendix~\ref{sec:controller}, we assume the sensor can also calculate the controller's state estimate $\hat{x}(t)$.

The state estimation error is defined as~\eqref{eq:state-estimation-error}, thus for $t \in (t_c^k,t_c^{k+1})$ we have
\begin{align}
    \label{nonlinearstateestii}
    \dot{z}=f(x(t),u(t),w(t))-f(\hat{x}(t),u(t),0).
\end{align}
A triggering occurs at time 
\begin{align}
\label{triggerinperiod}
t_s^{k}=k (\alpha+\gamma)
\end{align}
  and the sensor transmits a packet $p(t_s)$ of length $g(t_s)$ to the controller if
\begin{align}\label{eq:etsnonlinear}
  |z_1(t_s^{k})|\ge J,
\end{align}
where $J$ and $\alpha$ are non-negative real numbers, $\gamma$ is the upper bound on the channel delay, $k \in \mathbb{N}$, and $t_s^0=0$. We choose $g(t_s)$ such that after   decoding we have
\begin{align}
\label{jumpnonlineartgggg11!}
    |z(t_c^{k+})| \le J.
\end{align}
Clearly, the periodic event-triggering scheme~\eqref{triggerinperiod} and~\eqref{eq:etsnonlinear} does not  exhibit Zeno behavior, meaning that there cannot be infinitely many
 triggering events in a finite time interval. In fact, we have
\begin{align}
\label{emmifeijef2222!!!!!!!!!!}
	\Delta'_k =t_s^{k+1}-t_s^k \ge  \alpha+\gamma.
\end{align} 
\begin{assume}
\label{lishitz1}
The dynamic~\eqref{nonlinearsystem1} satisfies the Lipschitz property
\begin{align}
\label{lishitssequ}
    |f(x,u,w)-f(\hat{x},u,0)| \le L_x |x-\hat{x}|+ L_w |w|,
\end{align}
where $L_x >0$, $L_w >0$, and 
\begin{align}
\label{upperboundepsi11}
    |z(t)|=|x(t)-\hat{x}(t)| \le \Upsilon (\gamma).
\end{align}
Here for all $0\le \vartheta \le \gamma$, $\Upsilon(\vartheta)$ is defined as follows
\begin{align}
\label{upsilondefinition}
    \Upsilon(\vartheta) :=J e^{L_x(\alpha+\gamma+\vartheta)}+\frac{L_wM}{L_x} \left(e^{L_x(\alpha+\gamma+\vartheta)}-1\right).~~
\end{align}
\end{assume}
The reason for choosing the specific value for $\Upsilon(\gamma)$ in~\eqref{upperboundepsi11} will become clear by looking at the following Lemma.
If a triggering occurs at time $t_s^k$, we
define
%$\underline{t}^k$ be the infimum of time $t \in (t_s^{k-1},t_s^k]$ such that $|z(t)|=J$. 
\begin{align}
\label{tbarka1}
    \underline{t}^k = \mbox{inf}\left\{t \in (t_s^{k-1},t_s^k]~;~|z(t)|=J\right\}.
\end{align}
By continuity of $z$ during the inter-reception time, and using~\eqref{nonlinearstateestii} and~\eqref{jumpnonlineartgggg11!}, we see that $\underline{t}^k$ is well defined. This definition is used in the next Lemma. 
%The proof of this Lemma can be found in the 
%online 
%appendix.
%~\cite{supplementary2019}.
%By continuity if a triggering happens at time $t_s^k$, then there exists a time $\underline{t}^k \in (t_s^{k-1},t_s^k]$ such that $|z(\underline{t}^k)|=J$. 
%This point is helpful in the next Lemma.
\begin{lemma}\label{lem:uperboundznonlinear}
  Consider the plant-sensor-channel-controller model with plant dynamics~\eqref{nonlinearsystem1} satisfying Lipschitz property~\eqref{lishitssequ}, estimator dynamics~\eqref{nonlinearsystemestime}, triggering strategy~\eqref{triggerinperiod}, and~\eqref{eq:etsnonlinear}.
  %, and jump strategy~\eqref{eq:jumpst}. 
  Assume $|z(0)|=|x(0)-\hat{x}(0)|<J$ and~\eqref{jumpnonlineartgggg11!} occurs at all reception times $\{t^k_c\}_{k \in \mathbb{N}}$.
  %Assume the controller has enough information about $x(0)$ such that $|z(0)|<J$ and the packet size is large enough to ensure~\eqref{jumpnonlineartgggg11!} for all reception times $\{t^k_c\}_{k \in \mathbb{N}}$. %If a triggering occurs at time 
  Then for all time $t \in [\underline{t}^k,t_c^k)$, where $\vartheta=t-t_s^k$, we have 
 \begin{align}\label{definnotation11!!!22}
     &|z(t)| \le 
     \\
     &\Upsilon_{w}(\vartheta):=J e^{L_x(\alpha+\gamma+\vartheta)}+\frac{L_w|w|_t}{L_x} \left(e^{L_x(\alpha+\gamma+\vartheta)}-1\right).
 \end{align}
  %$ \le  \Upsilon(\vartheta)$,
  %if~\eqref{eq:jump-upp} occurs for all $k \in \mathbb{N}$ we have
  %$t_s \le t \le t_c$,
  %where .
  %and $\Upsilon$ is defined in~\eqref{upsilondefinition}.
  %\begin{align}\label{upponznonlinear}
  %  &|z(t)|  \le 
   %  \Upsilon(\vartheta) := 
%    \\\nonumber
 %   & J e^{L_x\left(b+\gamma+\vartheta\right)}+\frac{ML_w}{L_x} \left(e^{L_x\left(b+\gamma+\vartheta\right)}-1\right).
 % \end{align}
\end{lemma}
 \lemref{lem:uperboundznonlinear} has two important implications. First, if a triggering event does not occur at $t_s^k$ for all $t \in (t_s^{k-1},t_s^k]$ we have $|z(t)|\le J$, hence using~\eqref{jumpnonlineartgggg11!},
%If a triggering does not occur at time $t_s^k$ in the time interval from the last triggering time to $t_s^k$ we have $|z(t)|\le J$.
%from the last reception time to $t_0$ we have $z(t) \le J$ also 
under the assumptions of \lemref{lem:uperboundznonlinear} for all time $t\ge 0$ we have 
\begin{align}\label{22!!!444orkvve}
    |z(t)| \le  \Upsilon_{w}(\vartheta) \overset{(a)}{\le} \Upsilon_{w}(\gamma) \overset{(b)}{\le} \Upsilon(\gamma),
\end{align}
where $(a)$ follows from $\vartheta \le \gamma$, and $(b)$ follows from~\eqref{disturbupperbounnonlinearcase} and~\eqref{upsilondefinition}. 
Also, this last inequality explains why we defined the Lipschitz property as~\eqref{upperboundepsi11}. 
The second important implication of  \lemref{lem:uperboundznonlinear} is that for all $k \in \mathbb{N}$ we have
\begin{align}
    z(t_s^k) \in [- \Upsilon(0), \Upsilon(0)].
\end{align}

To construct the packet $p(t_s)$ of length $g(t_s)$, we 
uniformly quantize the interval $[- \Upsilon(0), \Upsilon(0)]$ into $2^{g(t_s)}$ equal intervals of size $2\gamma(0)/2^{g(t_s)}$.
Once the controller receives the packet, it determines the correct sub-interval and selects  its center point as the estimate of $z(t_s^k)$, which is represented by $\bar{z}(t_s)$.
In this case, we have
\begin{align}\label{kfejiw22223322!}
    |z(t_s)-\bar{z}(t_s)| \le \Upsilon(0)/2^{g(t_s)}.
\end{align}
By~\eqref{eq:state-estimation-error} we have $x(t_s)=z(t_s)+\hat{x}(t_s)$, thus using $\bar{z}(t_s)$ the controller can construct an estimate of $x(t_s)$ which we denote by $\bar{x}(t_s)$ as follows 
\begin{align}
\label{initialvalues}
   \bar{x}(t_s)=\bar{z}(t_s)+\hat{x}(t_s).
\end{align}
By~\eqref{kfejiw22223322!} we deduce that
\begin{align}\label{erroro1113!}
    |\bar{x}(t_s)-x(t_s)| \le \Upsilon(0)/2^{g(t_s)}.
\end{align}
For all $t \in [t_s,t_c]$ consider the differential equation
\begin{align}
\label{fejjefiij22222211!!}
    \dot{\bar{x}}=f(\bar{x}(t),u(t),0)
\end{align}
with initial condition $\bar{x}(t_s)$ given in~\eqref{initialvalues},
and let its solution at time $t_c$ be equal to $\hat{x}(t_c^+)$, namely  
\begin{align}\label{estimNonlin}
   \hat{x}(t_c^+)=\bar{x}(t_s)+\int_{t_s}^{t_c}f(\bar{x}(t),u(t),0).
\end{align}
We use the above quantization policy to find a sufficient packet size in the next Theorem.
%, and its proof is relegated to the appendix.
%can be found in~\cite{supplementary2019}.
\begin{theorem}
\label{Thmmainpacketnolin!!!}
Consider the plant-sensor-channel-controller model with plant dynamics~\eqref{nonlinearsystem1} with Lipschitz property~\eqref{lishitssequ}, estimator dynamics~\eqref{nonlinearsystemestime}, triggering strategy~\eqref{triggerinperiod}, and~\eqref{eq:etsnonlinear}.
  %, and jump strategy~\eqref{eq:jumpst}. 
  %Assume the controller has enough information about $x(0)$ such that $|z(0)|<J$. 
   Assume $|z(0)|=|x(0)-\hat{x}(0)|<J$,
  then there exists a
quantization policy that achieves~\eqref{jumpnonlineartgggg11!} for all reception times $\{t^k_c\}_{k \in \mathbb{N}}$ with any packet size
\begin{small}
\begin{align}
\label{nifekn2!!!!!2}
  g(t_s) \ge  \max\left\{0,\log\left(\frac{\Upsilon(0)e^{L_x \gamma}}{J-\frac{L_wM}{L_x}\left(e^{L_x \gamma}-1\right)}\right) \right\},
  \end{align}
  \end{small}
  provided
  \begin{align}
  \label{conditiontocalculatee3444}
J \ge \frac{L_wM}{L_x}\left(e^{L_x \gamma}-1\right).
\end{align}
\end{theorem}
In the next assumption we restrict the class of nonlinear systems.
\begin{assume}
\label{ISS1}
There
exists a control policy $u(t)=\mathfrak{U}(\hat{x})=\mathfrak{U}(x-z)$ which renders the dynamics~\eqref{nonlinearsystem1} ($\dot{x}=f(x,\mathfrak{U}(x-z),w)$)
ISS with respect to $z(t)$ and $w(t)$, that is, there exists $\beta' \in \mathcal{KL}$, $\Pi' \in \mathcal{K}_\infty(0)$, and $\psi' \in \mathcal{K}_\infty(0)$  such that for all $ t\ge 0$
\begin{align}
    |x(t)|\le \beta'\left(|x(0)|,t\right)+\Pi'\left(|z|_t\right)+\psi'\left(|w|_t\right).
    %\sup_{s \in [0,t]} |z(s)|
    %\sup_{s \in [0,t]} |w(s)|
\end{align}
\end{assume}
%The proof of the following Corollary is in the appendix.
\begin{corollary}
\label{inefiie!!!22222}
Under the assumptions of \thmref{Thmmainpacketnolin!!!} and \asumref{ISS1}  for any packet size lower bounded as~\eqref{nifekn2!!!!!2} there exists a control policy which renders the dynamics~\eqref{nonlinearsystem1} ISpS.
%as stated in Defintion
\end{corollary}
Using~\eqref{emmifeijef2222!!!!!!!!!!} the triggering rate, the frequency at which triggering occurs, is trivially upper bounded by $(\alpha+\gamma)^{-1}$. As a result, under assumptions of Corollary~\ref{inefiie!!!22222}  we deduce that for any information
transmission rate~\eqref{ratemo232oe4e}
%\begin{small}
  \begin{align}\label{inftranratenonlinear2}
R_s \ge 
\frac{1}{\alpha+\gamma}\max\left\{0,\log\left(\frac{\Upsilon(0)e^{L_x \gamma}}{J-\frac{L_wM}{L_x}\left(e^{L_x \gamma}-1\right)}\right) \right\},~~\,
  \end{align}
  %\end{small}
  there exists a control law that renders the dynamic~\eqref{nonlinearsystem1} ISpS.
  
The interested reader can find some additional remarks and simulations for the nonlinear systems in Appendix~\ref{sec:remnon} and~\ref{sec:sim}.
\section{Future work}\label{sec:conc}
%We presented an experimental validation of an event-triggered policy for a continuous linear time-invariant system over a digital channel with random bounded delay while utilizing timing information is an state dependent fashion.
%by encoding both state and timing information in one packet. 
%We built a laboratory-scale inverted pendulum using off-the-shelf components and implemented the proposed control scheme on it. The controller is tested in various scenarios for different values of channel delay upper bound. The results verify the increase in packet size and sufficient transmission rate for stabilization as the channel delay increases. Finally, under appropriate assumptions we also provided  an encoding-decoding scheme to achieve input-to-state practically stability (ISpS) for nonlinear continuous-time systems.
On the theoretical side, future work will explore the theory and implementation of   multivariate nonlinear system with  uncertainty in its parameters. On the practical validation side, we also plan to test the proposed nonlinear scheme on our inverted pendulum prototype.
%and with multiple unstable modes.
%
%\newpage
%\newpage
%\textbf{Theory and implementation of event-triggered stabilization over digital channels (supplementary materials)}
%
%M. J. Khojasteh, M. Hedayatpour, M. Franceschetti
%\appendix
%\subsection{Proof of Lemma \ref{lem:uperboundznonlinear}}
%\subsection{Proof of Theorem \ref{Thmmainpacketnolin!!!}}
%\subsection{Proof of Corollary \ref{inefiie!!!22222}}
\bibliography{mybib} 

% Generated by IEEEtran.bst, version: 1.14 (2015/08/26)
\begin{thebibliography}{10}
\providecommand{\url}[1]{#1}
\csname url@samestyle\endcsname
\providecommand{\newblock}{\relax}
\providecommand{\bibinfo}[2]{#2}
\providecommand{\BIBentrySTDinterwordspacing}{\spaceskip=0pt\relax}
\providecommand{\BIBentryALTinterwordstretchfactor}{4}
\providecommand{\BIBentryALTinterwordspacing}{\spaceskip=\fontdimen2\font plus
\BIBentryALTinterwordstretchfactor\fontdimen3\font minus
  \fontdimen4\font\relax}
\providecommand{\BIBforeignlanguage}[2]{{%
\expandafter\ifx\csname l@#1\endcsname\relax
\typeout{** WARNING: IEEEtran.bst: No hyphenation pattern has been}%
\typeout{** loaded for the language `#1'. Using the pattern for}%
\typeout{** the default language instead.}%
\else
\language=\csname l@#1\endcsname
\fi
#2}}
\providecommand{\BIBdecl}{\relax}
\BIBdecl

\bibitem{Tabuada}
P.~Tabuada, ``Event-triggered real-time scheduling of stabilizing control
  tasks,'' \emph{IEEE Tran. Auto. Cont.}, vol.~52, no.~9, pp. 1680--1685, 2007.

\bibitem{wang2011event}
X.~Wang and M.~D. Lemmon, ``Event-triggering in distributed networked control
  systems,'' \emph{IEEE Tran. Auto. Cont.}, vol.~56, no.~3, pp. 586--601, 2011.

\bibitem{WPMHH-KHJ-PT:12}
W.~P. M.~H. Heemels, K.~H. Johansson, and P.~Tabuada, ``An introduction to
  event-triggered and self-triggered control,'' in \emph{Proc. IEEE Conf.
  Decis. Cont.}, 2012, pp. 3270--3285.

\bibitem{pearson2017control}
J.~Pearson, J.~P. Hespanha, and D.~Liberzon, ``Control with minimal
  cost-per-symbol encoding and quasi-optimality of event-based encoders,''
  \emph{IEEE Tran. Auto. Cont.}, vol.~62, no.~5, pp. 2286--2301, 2017.

\bibitem{tallapragada2013event}
P.~Tallapragada and N.~Chopra, ``On event triggered tracking for nonlinear
  systems,'' \emph{IEEE Tran. Auto. Cont.}, vol.~58, no.~9, pp. 2343--2348,
  2013.

\bibitem{postoyan2015framework}
R.~Postoyan, P.~Tabuada, D.~Ne{\v{s}}i{\'c}, and A.~Anta, ``A framework for the
  event-triggered stabilization of nonlinear systems,'' \emph{IEEE Tran. Auto.
  Cont.}, vol.~60, no.~4, pp. 982--996, 2015.

\bibitem{girard2015dynamic}
A.~Girard, ``Dynamic triggering mechanisms for event-triggered control,''
  \emph{IEEE Tran. Auto. Cont.}, vol.~60, no.~7, pp. 1992--1997, 2015.

\bibitem{yildiz2019event}
H.~Yildiz, Y.~Su, A.~Khina, and B.~Hassibi, ``Event-triggered stochastic
  control via constrained quantization,'' in \emph{Proc. IEEE Data Comp. Conf.
  (DCC)}, 2019, pp. 612--612.

\bibitem{linsenmayer2018containability}
S.~Linsenmayer, H.~Ishii, and F.~Allg{\"o}wer, ``Containability with
  event-based sampling for scalar systems with time-varying delay and
  uncertainty,'' \emph{IEEE Cont. Sys. Let.}, vol.~2, no.~4, pp. 725--730,
  2018.

\bibitem{seuret2016lq}
A.~Seuret, C.~Prieur, S.~Tarbouriech, and L.~Zaccarian, ``{L}{Q}-based
  event-triggered controller co-design for saturated linear systems,''
  \emph{Automatica}, vol.~74, pp. 47--54, 2016.

\bibitem{khashooei2018consistent}
B.~A. Khashooei, D.~J. Antunes, and W.~Heemels, ``A consistent threshold-based
  policy for event-triggered control,'' \emph{IEEE Cont. Sys. Let.}, vol.~2,
  no.~3, 2018.

\bibitem{khojasteh2017time}
M.~J. Khojasteh, P.~Tallapragada, J.~Cort\'{e}s, and M.~Franceschetti,
  ``Time-triggering versus event-triggering control over communication
  channels,'' in \emph{Proc. IEEE Conf. Decis. Cont.}, 2017, pp. 5432--5437.

\bibitem{khojasteh2018stabilizing}
M.~J. Khojasteh, M.~Franceschetti, and G.~Ranade, ``Stabilizing a linear system
  using phone calls,'' in \emph{Proc. Euro. Cont. Conf.}\hskip 1em plus 0.5em
  minus 0.4em\relax IEEE, 2019, pp. 2856--2861.

\bibitem{OurJournal1}
M.~J. Khojasteh, P.~Tallapragada, J.~Cort{\'e}s, and M.~Franceschetti, ``The
  value of timing information in event-triggered control,'' \emph{IEEE Tran.
  Auto. Cont.}, 2020.

\bibitem{Nair}
B.~G.~N. Nair, F.~Fagnani, S.~Zampieri, and R.~J. Evans, ``Feedback control
  under data rate constraints: An overview,'' \emph{Proceedings of the IEEE},
  vol.~95, no.~1, pp. 108--137, 2007.

\bibitem{Mitter}
S.~Tatikonda and S.~Mitter, ``Control under communication constraints,''
  \emph{IEEE Tran. Auto. Cont.}, vol.~49, no.~7, pp. 1056--1068, 2004.

\bibitem{khina2019control}
A.~Khina, E.~R. Garding, G.~M. Pettersson, V.~Kostina, and B.~Hassibi,
  ``Control over gaussian channels with and without source-channel
  separation,'' \emph{IEEE Tran. Auto. Cont.}, 2019.

\bibitem{kostina2016rate11}
V.~Kostina and B.~Hassibi, ``Rate-cost tradeoffs in control,'' \emph{IEEE Tran.
  Auto. Cont.}, 2019.

\bibitem{Massimo}
M.~Franceschetti and P.~Minero, ``Elements of information theory for networked
  control systems,'' in \emph{Info. and Cont. in Net.}\hskip 1em plus 0.5em
  minus 0.4em\relax Springer, 2014, pp. 3--37.

\bibitem{martins2006feedback}
N.~C. Martins, M.~A. Dahleh, and N.~Elia, ``Feedback stabilization of uncertain
  systems in the presence of a direct link,'' \emph{IEEE Tran. Auto. Cont.},
  vol.~51, no.~3, pp. 438--447, 2006.

\bibitem{hespanha2002towards}
J.~Hespanha, A.~Ortega, and L.~Vasudevan, ``Towards the control of linear
  systems with minimum bit-rate,'' in \emph{Proc. 15th Int. Symp. Math. Theory
  Netw. Syst.}, 2002.

\bibitem{princeton_paper}
M.~J. Khojasteh, M.~Hedayatpour, J.~Cort\'{e}s, and M.~Franceschetti,
  ``Event-triggered stabilization of disturbed linear systems over digital
  channels,'' in \emph{Ann. Conf. on Inf. Sci. and Sys.}\hskip 1em plus 0.5em
  minus 0.4em\relax IEEE, 2018, pp. 1--6.

\bibitem{khojasteh2018event21122}
M.~J. Khojasteh, M.~Hedayatpour, J.~Cortes, and M.~Franceschetti,
  ``Event-triggered stabilization over digital channels of linear systems with
  disturbances,'' \emph{arXiv preprint arXiv:1805.01969}, 2018.

\bibitem{jiang1994small}
Z.-P. Jiang, A.~R. Teel, and L.~Praly, ``Small-gain theorem for {I}{S}{S}
  systems and applications,'' \emph{Math. of Cont., Sig. and Sys.}, vol.~7,
  no.~2, pp. 95--120, 1994.

\bibitem{sharon2012input}
Y.~Sharon and D.~Liberzon, ``Input to state stabilizing controller for systems
  with coarse quantization,'' \emph{IEEE Tran. Auto. Cont.}, vol.~57, no.~4,
  pp. 830--844, 2012.

\bibitem{Topological}
G.~N. Nair, R.~J. Evans, I.~M. Mareels, and W.~Moran, ``Topological feedback
  entropy and nonlinear stabilization,'' \emph{IEEE Tran. Auto. Cont.},
  vol.~49, no.~9, pp. 1585--1597, 2004.

\bibitem{liberzon2005stabilization}
D.~Liberzon and J.~P. Hespanha, ``Stabilization of nonlinear systems with
  limited information feedback,'' \emph{IEEE Tran. Auto. Cont.}, vol.~50,
  no.~6, pp. 910--915, 2005.

\bibitem{de2005n}
C.~De~Persis, ``n-bit stabilization of n-dimensional nonlinear systems in
  feedforward form,'' \emph{IEEE Tran. Auto. Cont.}, vol.~50, no.~3, pp.
  299--311, 2005.

\bibitem{liberzon2003stabilization}
D.~Liberzon, ``On stabilization of linear systems with limited information,''
  \emph{IEEE Tran. Auto. Cont.}, vol.~48, no.~2, pp. 304--307, 2003.

\bibitem{sanjaroon2018estimation}
V.~Sanjaroon, A.~Farhadi, A.~S. Motahari, and B.~H. Khalaj, ``Estimation of
  nonlinear dynamic systems over communication channels,'' \emph{IEEE Tran.
  Auto. Cont.}, vol.~63, no.~9, pp. 3024--3031, 2018.

\bibitem{sontag2008input}
E.~D. Sontag, ``Input to state stability: Basic concepts and results,'' in
  \emph{Nonlin. and Opt. Cont. The.}\hskip 1em plus 0.5em minus 0.4em\relax
  Springer, 2008, pp. 163--220.

\bibitem{dou2018sufficient}
R.~Dou, J.~Chen, F.~Li, S.~Wang, and Q.~Ling, ``A sufficient bit rate condition
  for stabilizing a scalar continuous-time nonlinear system with bounded
  processing delay,'' in \emph{37th Chin. Cont. Conf.}\hskip 1em plus 0.5em
  minus 0.4em\relax IEEE, 2018, pp. 6265--6270.

\bibitem{tanwani2017stabilization}
A.~Tanwani and A.~Teel, ``Stabilization with event-driven controllers over a
  digital communication channel with random transmissions,'' in \emph{Proc.
  IEEE Conf. Decis. Cont.}, 2017, pp. 6063--6068.

\bibitem{music2018design}
Z.~Music, F.~Molinari, S.~Gallenm{\"u}ller, O.~Ayan, S.~Zoppi, W.~Kellerer,
  G.~Carle, T.~Seel, and J.~Raisch, ``Design of a networked controller for a
  two-wheeled inverted pendulum robot,'' \emph{arXiv preprint
  arXiv:1812.03071}, 2018.

\bibitem{nair2004stabilizability}
G.~N. Nair and R.~J. Evans, ``Stabilizability of stochastic linear systems with
  finite feedback data rates,'' \emph{SIAM Jour. on Cont. and Opt.}, vol.~43,
  no.~2, pp. 413--436, 2004.

\bibitem{tatikonda2004control}
S.~Tatikonda and S.~Mitter, ``Control over noisy channels,'' \emph{IEEE Tran.
  Auto. Cont.}, vol.~49, no.~7, pp. 1196--1201, 2004.

\bibitem{heemels2013periodic}
W.~H. Heemels, M.~Donkers, and A.~R. Teel, ``Periodic event-triggered control
  for linear systems,'' \emph{IEEE Tran. Auto. Cont.}, vol.~58, no.~4, pp.
  847--861, 2013.

\bibitem{de2004stabilizability}
C.~De~Persis and A.~Isidori, ``Stabilizability by state feedback implies
  stabilizability by encoded state feedback,'' \emph{Sys. \& cont. let.},
  vol.~53, no. 3-4, pp. 249--258, 2004.

\bibitem{hespanha2008lyapunov}
J.~P. Hespanha, D.~Liberzon, and A.~R. Teel, ``Lyapunov conditions for
  input-to-state stability of impulsive systems,'' \emph{Automatica}, vol.~44,
  no.~11, pp. 2735--2744, 2008.

\end{thebibliography}
\bibliographystyle{IEEEtran}
%\newpage
%\newpage
%\onecolumn
%\twocolumn
%\newpage
\appendix
\subsection{Notation}
Throughout the paper, we use the following notation. We represent the set of real, non-negative real, and natural numbers by $\real$, $\real_+$, and $\integers$, respectively.
Base $2$ and natural logarithms are  represented by $\log$ and $\ln$ respectively. Vectors are represented by boldface italic letters and matrices are represented by regular and capital boldface letters. We use regular lowercase letters to represent scalars. To represent an element of a vector, we use the vector name accompanied by the element's index as its subscript. 
For a function $f : \real \rightarrow \real^n$ and $t \in \real$, the right-hand limit of $f$ at $t$ or $\lim_{s \rightarrow t^+} f(s)$ is represented by $f(t^+)$. 
Also, the nearest integer less (resp. greater) than or equal to $x$ is represented by $\floor{x}$ (resp. $\ceil{x}$). The remainder after division of $x$ by~$y$ is indicated by the modulo function as $\modulo(x,y)$ and $\text{sign}(x)$ returns the sign of~$x$. For a scalar continuous-time signal $w(t)$, we define
%\begin{align}
    $|w|_t=\sup_{s \in [0,t]} |w_1(s)|$.
%\end{align}
Finally, to formulate the stability properties,  for non-negative constants $d$ and $d'$ we define
\begin{align}
    &\mathcal{K}(d):= 
    \{f:\real_{\ge 0} \rightarrow \real_{\ge 0} | f~\mbox{continuous,}
     \\
        &~~~~~~~~~~~~~~~~
    \mbox{strictly
increasing, and}~f(0)=d\},
\\
&\mathcal{K_\infty}(d):= 
    \{f \in \mathcal{K}(d)| f~\mbox{unbounded}\},
    \\
         &\mathcal{K}_\infty^2(0,d'):= 
    \{f:\real_{\ge 0} \times \real_{\ge 0} \rightarrow \real_{\ge 0} | \\
    &~~~~~\forall t \ge 0, f(.,t) \in \mathcal{K}_\infty(0),~\mbox{and}~\forall r> 0~f(r,.) \in \mathcal{K}_\infty(d')\}
    \\
    &\mathcal{L}:= 
    \{f:\real_{\ge 0} \rightarrow \real_{\ge 0} | f~\mbox{continuous,}\\
    &~~~~~\mbox{strictly
    decreasing, and}~\lim_{s \rightarrow \infty} f(s)=0\},
    \\
     &\mathcal{KL}:= 
    \{f:\real_{\ge 0} \times \real_{\ge 0} \rightarrow \real_{\ge 0} | f~\mbox{continuous,}\\
    &~~~~~\forall t \ge 0, f(.,t) \in \mathcal{K}(0),~\mbox{and}~\forall r>0~f(r,.) \in \mathcal{L}\}.
\end{align}
\subsection{Event-triggering control design}\label{sec:controller}
We start with a brief description of the event-triggered control approach  in our previous work~\cite{princeton_paper,khojasteh2018event21122} that achieves ISpS for the first coordinate of the dynamics~\eqref{AAmj}. We have 
\begin{align}\label{syscon}
   \dot{x}_1=\lambda_1 x_1(t)+b_1u(t)+w_1(t).
\end{align}
%The  mode of the system corresponding to the unstable pole (the first element of $\tilde{\pmb{x}}$), can be described by a scalar, continuous-time, linear time-invariant model. Defining $x=3.6940\phi + 0.5046\dot{\phi}$, $\lambda=7.3198$, $b=0.2523$ and $w(t)\le 0.0470$, we can write the following: 
%where $x(t) \in \real$ and $u(t) \in \real$ for $t \in [0,\infty)$ are the plant state and control input, respectively, and $w(t)  \in \real$ represents the plant disturbance. The latter is upper bounded as: 
%\begin{align}\label{noiseupp}
%|w(t)|\le M,
%\end{align}
%where $M$ is a  nonnegative real number. In~\eqref{syscon}, $A$ is a positive real number (i.e., the plant is unstable),  $B \in \real$ and the initial condition $x(0)$ is bounded. 
%\subsection{Networked control system }
At the controller, the estimated state is represented by $\hat{x}_1$ and evolves during the inter-reception times as
\begin{align}\label{sysest}
  \dot{\hat{x}}_1(t)=\lambda_1\hat{x}_1(t)+b_1u(t), \quad t \in (t_c^k,t_c^{k+1}),
\end{align}
starting at $\hat{x}_1(t_c^{k+})$, as the estimate of the state at the controller is updated with the information received up to time $t_c^{k+}$. 
%This estimate of state 
$\hat{x}_1(t_c^{k+})$ is found by decoding the received packet,
%quantized version of the state 
as explained later in this section.
We 
%let $\hat{x}_1(0)=\hat{x}_0$ and
assume that the sensor can  also construct 
the estimate $\hat{x}_1(t)$ generated by the controller.
\begin{remark} {\rm 
As shown in~\cite{princeton_paper,khojasteh2018event21122}, if the sensor has causal knowledge of the delay in the
communication channel, it can use $\hat{x}(0)$  to compute  $\hat{x}(t)$ at all times $t$. This causal knowledge  can be obtained without assuming an additional communication channel in the feedback loop via ``acknowledgment through the control input''~\cite{tatikonda2004control}.
  }
  \oprocend
\end{remark}

%starting from $\hat{x}_1(t_c^{k+})$which represents the state estimate of the controller with the information received up to time $t_c^k$ (The exact way to construct $\hat{x}_1(t_c^{k+})$ is explained later in this section). We also set  $\hat{x}_1(0)=\hat{x}_0$. 
%We assume that the sensor knows $\hat{x}_0$ and has knowledge of the times the actuator performs the control action. This is to ensure that the sensor can also compute $\hat{x}(t)$ for all time $t$. 
%As proved in~\cite{princeton_paper} provided  causal knowledge of the delay in the communication channel sensor can track the value of $\hat{x}(t)$ for all time $t$.
%In practice, this corresponds to assuming an   instantaneous acknowledgment   from the actuator to the sensor via the control input, as discussed in~\cite{sahai2006necessity,ling2017bit}. 
%To obtain such causal knowledge, one can monitor the output of the actuator provided that the control input changes at each reception time.  In case the sensor has only access to the plant state, one can use a narrowband signal in the control input to excite a specific frequency of the state,  that can signal the time at which the control action has been applied. 
For the unstable coordinate in~\eqref{AAmj}, the \emph{state estimation error} is defined as
\begin{align}\label{eq:state-estimation-error}
z_1(t)=x_1(t)-\hat{x}_1(t).
\end{align}
%where $z_1(0)=x_1(0)-\hat{x}_0$. 
This error is used by the sensor to determine the triggering events. 
%We rely on this error to determine when a triggering event occurs in our controller design.

We define the triggering events as follows: for $J \ge 0$, a triggering occurs, and the sensor transmits a packet to the controller at time $t_s^{k+1} \ge 0$ when
\begin{align}\label{eq:ets}
  |z_1(t_s^{k+1})|=J,
\end{align}
where $t_c^k \le t_s^{k+1}$ for $k \in \mathbb{N}$ and $t_s^1 \ge 0$. Thus, a new transmission occurs only if the previous packets have been already delivered to the controller.
%Also, we assume that a new transmission occurs only if the previous packets have been already delivered to the controller. %This is how we are encoding timing information as the controller knows that, after receiving the packet at time $t_c^k$, the triggering event has already occurred at time $t_s^k \in [t_c^k-\gamma,t_c^k]$ and~\eqref{eq:ets} is satisfied. 

Using~\eqref{eq:ets}, at each triggering event, the sensor transmits the packet $p(t_s)$ of size $g(t_s)$ bits to the controller which contains the data payload %plus encoded timing information.
and carries the timing information.
The details of our quantization policy to encode data payload into packet $p(t_s)$ is discussed in our previous work~\cite{princeton_paper,khojasteh2018event21122}.
Using the data payload and the timing of the packet $p(t_s)$ the controller estimates $z_1(t_c)$ as follows
%Once the packet is complete, it is transmitted to the controller, where it is decoded and the center point of the smallest sub-interval is selected as the best estimate of~$t_s$ which we denote by $q(t_s)$. Then the controller estimates $z_1(t_c)$ as 
\begin{align}\label{ctrlapp1}
   \bar{z}_1(t_c)=\text{sign}(z_1(t_s)) J e^{\lambda_1(t_c-q(t_s))},
\end{align}
where $q(t_s)$ is the best estimate of $t_s$ constructed at the controller after reception of the packet $p(t_s)$.

To update the estimate of the state after decoding the packet, we define the following \emph{jump strategy}  
%Using this estimation, the controller updates the state estimate using the \emph{jump strategy}, 
\begin{align}
 \hat{x}_1(t_c^+)=\bar{z}_1(t_c)+\hat{x}_1(t_c).
 \label{eq:jumpst}
\end{align}
For a given design parameter $0<\rho_0<1$,   the packet size  
 \begin{align}\label{packet_size}
   g(t_s) = \max\left\{1,\left\lceil1+\log
       \frac{\lambda_1b\gamma}{\ln(1+\frac{\rho_0-(M/J\lambda_1)(e^{\lambda_1\gamma}-1)}{e^{\lambda_1\gamma}})}\right\rceil\right\}, 
\end{align}
ensures that
\begin{align}
|z_1(t_c^{k+})| =|z_1(t_c^k)-\bar{z}_1(t_c^k)| \le \rho_0 J,
   \label{eq:jump-upp}
\end{align}
at all reception times $\{t_c^k\}_{k \in \mathbb{N}}$   inside the closed interval $[t_s^k,t_s^k+\gamma]$, provided that $J> \frac{M}{\lambda_1\rho_0}(e^{\lambda_1\gamma}-1)$ and
$|z_1(0)|\le J$.
It then follows that   for all   $t$ we have
  \begin{align}\label{upponz}
    |z_1(t)|\le J e^{\lambda_1\gamma}+\frac{M}{\lambda_1} \left(e^{\lambda_1\gamma}-1\right).
  \end{align}
The proof and derivation of~\eqref{packet_size},~\eqref{eq:jump-upp}, and~\eqref{upponz} can be found in our previous work~\cite{princeton_paper,khojasteh2018event21122}. 

We set the control input to be
$u(t)=-\textbf{K}\hat{\pmb{x}}(t)$. 
In our example, we have $\textbf{K}=(225,11)$, and \textbf{K} is chosen such that $(\textbf{A}-\textbf{B}\textbf{K})$ is Hurwitz. Here, $\hat{x}_1(t)$ is generated based on~\eqref{sysest} and jump strategy~\eqref{eq:jumpst}. 
Since $\lambda_2 <0$, the controller does not need any update from the sensor to construct $\hat{x}_2(t)$. Therefore, as mentioned in~\cite{princeton_paper}, using~\eqref{upponz}, one can prove that the plant~\eqref{sys_dyn1} is ISpS. We can then conclude that as long as the linear approximation holds,  the nonlinear plant~\eqref{nonlinear_sys} is also ISpS.  

In~\cite{princeton_paper,khojasteh2018event21122} we have also shown that the proposed event-triggered scheme  does not  exhibit ``Zeno behavior,'' meaning that there cannot be infinitely many
 triggering events in a finite time interval. In fact,   the time between consecutive triggers is uniformly lower bounded as follow
 \begin{align}
	\Delta'_k =t_s^{k+1}-t_s^k \ge \frac{1}{\lambda_1} \ln \Big(\frac{J+\frac{M}{\lambda_1}}{\rho_0J+\frac{M}{\lambda_1}} \Big).
\end{align} 
%\section{Description of the prototype and system architecture}
\subsection{Proof of Lemma \ref{lem:uperboundznonlinear}}
\begin{proof}
%Since a triggering occurred at time $t_s^k$, $|z(t)|=J$ for some time $t_0 \in (t_s^{k-1},t_s^k]$. 
%Consequently, 
For all time $t \in [\underline{t}^k,t_c^k)$ the state estimation error evolves according to~\eqref{nonlinearstateestii} with the initial condition $z(\underline{t}^k)=J$, where $\underline{t}^k$ is defined as~\eqref{tbarka1}. Thus,
%starting form $t_0$ to $t_c^k$ 
for all $t \in [\underline{t}^k,t_c^k)$
\begin{subequations}
\begin{align}
    &|z(t)| \le Je^{L_x (t-\underline{t}^k)}+L_w\int_{\underline{t}^k}^{t}|w(t)e^{L_x(t-\underline{t}^k)}|dt 
    \label{1equaitionlong12!}
    \\ 
    &\le Je^{L_x (t-\underline{t}^k)}+\frac{L_w|w|_t}{L_x}\left(e^{L_x (t-\underline{t}^k)}-1\right)
    \\ 
    &=Je^{L_x (t-t_s+t_s-\underline{t}^k)}+\frac{L_w|w|_t}{L_x}\left(e^{L_x (t-t_s+t_s-\underline{t}^k)}-1\right)
    \\ 
    &\le Je^{L_x (\vartheta+\alpha+\gamma)}+\frac{L_w|w|_t}{L_x}\left(e^{L_x (\vartheta+\alpha+\gamma)}-1\right),
    \label{2equaitionlong12!}
\end{align}
\end{subequations}
where~\eqref{1equaitionlong12!} follows from the Lipschitz property~\eqref{lishitssequ} and Gronwall-Bellman inequality, as $\underline{t}^k \in (t_s^{k-1},t_s^k]$ we have $t_s^k-\underline{t}^k \le \alpha+\gamma$ and~\eqref{2equaitionlong12!} follows.
\end{proof}
\subsection{Proof of Theorem \ref{Thmmainpacketnolin!!!}}
\begin{proof}
For all $t \in [t_s,t_c]$ we have
\begin{subequations}
\label{0inmany}
\begin{align}
    &|x(t)-\bar{x}(t)|=
    \left|x(t_s)-\bar{x}(t_s)\right|+
    \\
    &\left|\int_{t_s}^{t} f(x,u,w) dt -\int_{t_s}^{t} f(\bar{x},u,0) dt\right| \le
    \label{1inmany}
    \\
    &|x(t_s)-\bar{x}(t_s)|+\int_{t_s}^{t}\left(L_x|x-\bar{x}|+L_w|w|\right) dt \le
    \label{2inmany}
    \\
    &|x(t_s)-\bar{x}(t_s)|e^{L_x (t-t_s)}+
    \\
    &~~~~~~~~~~~~~~L_w\int_{t_s}^{t}|w(t)e^{L_x(t-t_s)}|dt \le
    \label{3inmany}
    \\
    &|x(t_s)-\bar{x}(t_s)|e^{L_x (t-t_s)}+\frac{L_wM}{L_x}\left(e^{L_x (t-t_s)}-1\right)
    \label{4inmany}
\end{align}
\end{subequations}
where we used~\eqref{nonlinearsystem1} and~\eqref{fejjefiij22222211!!} along the triangle inequality to arrive at~\eqref{1inmany},~\eqref{2inmany} follows from Lipschitz property~\eqref{lishitssequ}, and~\eqref{3inmany} follows from solving the linear differential equation $\dot{x}(t)-\dot{\bar{x}}(t)=L_x(x-\bar{x})+L_ww$ with initial condition $x(t_s)-\bar{x}(t_s)$ (see Gronwall-Bellman inequality), and~\eqref{4inmany} follows from~\eqref{disturbupperbounnonlinearcase}.

Using~\eqref{gammma},~\eqref{erroro1113!},~\eqref{estimNonlin} and~\eqref{0inmany} we deduce
\begin{align}
    &|z(t_c^+)|=|x(t_c)-\hat{x}(t_c^+)|\le 
    \\
    &~~~~~~~~~~~~~~~~~~~~\frac{\Upsilon(0)}{2^{g(t_s)}}e^{L_x \gamma}+\frac{L_wM}{L_x}\left(e^{L_x \gamma}-1\right).
\end{align}
Consequently,
\begin{align}\label{femeif2222344411!!!!}
  \frac{\Upsilon(0)}{2^{g(t_s)}}e^{L_x \gamma}+\frac{L_wM}{L_x}\left(e^{L_x \gamma}-1\right) \le J
\end{align}
suffices to ensure~\eqref{jumpnonlineartgggg11!} at all reception time. Using~\eqref{conditiontocalculatee3444},~\eqref{femeif2222344411!!!!} is equivalent to
%\begin{align}
%  \frac{\Upsilon(0)}{2^{g(t_s)}}e^{L_x \gamma} \le J-\frac{L_wM}{L_x}\left(e^{L_x \gamma}-1\right)
%\end{align}
\begin{align}
  2^{g(t_s)} \ge \frac{\Upsilon(0)e^{L_x \gamma}}{J-\frac{L_wM}{L_x}\left(e^{L_x \gamma}-1\right)}.  
\end{align}
%\begin{align}
%  g(t_s) \ge  \log\left(\frac{\Upsilon(0)e^{L_x \gamma}}{J-\frac{L_wM}{L_x}\left(e^{L_x \gamma}-1\right)}\right) 
%  \end{align}
%  maximum 1 the result follows.
The result now follows by noticing the packet size should be no-negative.
\end{proof}
\subsection{Proof of Corollary \ref{inefiie!!!22222}}
\begin{proof}
\thmref{Thmmainpacketnolin!!!} states that with any packet size lower bounded  as~\eqref{nifekn2!!!!!2} there exists a
quantization policy that achieves~\eqref{jumpnonlineartgggg11!} for all reception times $\{t^k_c\}_{k \in \mathbb{N}}$. Thus using \lemref{lem:uperboundznonlinear} and~\eqref{22!!!444orkvve} we deduce for all time $t \ge 0$ we have
\begin{align}
|z(t)| \le  \Upsilon_{w}(\gamma),
\end{align}
where $\Upsilon_{w}(\gamma)$ is defined as~\eqref{definnotation11!!!22}. Consequently, for all time $t \ge 0$, $|z(t)|$ is upper bounded by summation of a $\mathcal{K}_\infty(d)$ function of $\gamma$ with $d=J e^{L_x \alpha}$ and a $\mathcal{K}_\infty^2(0,d')$  function of $|w|_t$ and $\gamma$ with $d'=(e^{L_x \alpha}-1)L_w M/L_x$. Therefore, using \asumref{ISS1} the result follows.
\end{proof}
\subsection{Additional remarks about nonlinear systems}\label{sec:remnon}
\begin{figure*}[t]
\centering
\begin{tabular}{c c c}
	% \scriptsize{$M=0$, $\gamma=0.1$ sec, $g(t_s)=1$ bit} & 
	 \scriptsize{$M=0.1$, $\gamma=0.1$ sec, $g(t_s)=3$ bit} &
	 \scriptsize{$M=0.1$, $\gamma=0.99$ sec, $g(t_s)=15$ bit} \\
    \includegraphics[width=60mm]{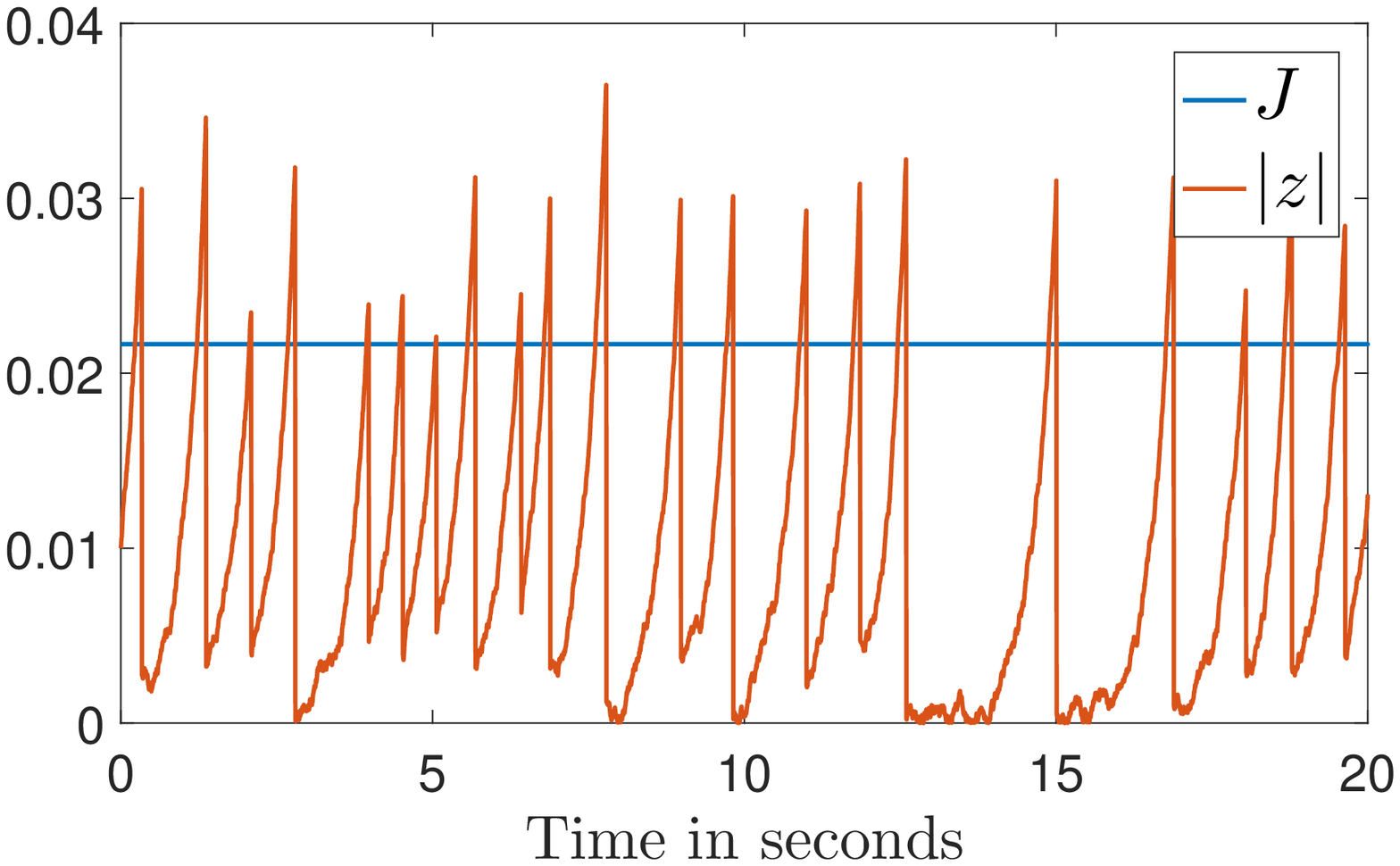} & 
    \includegraphics[width=60mm]{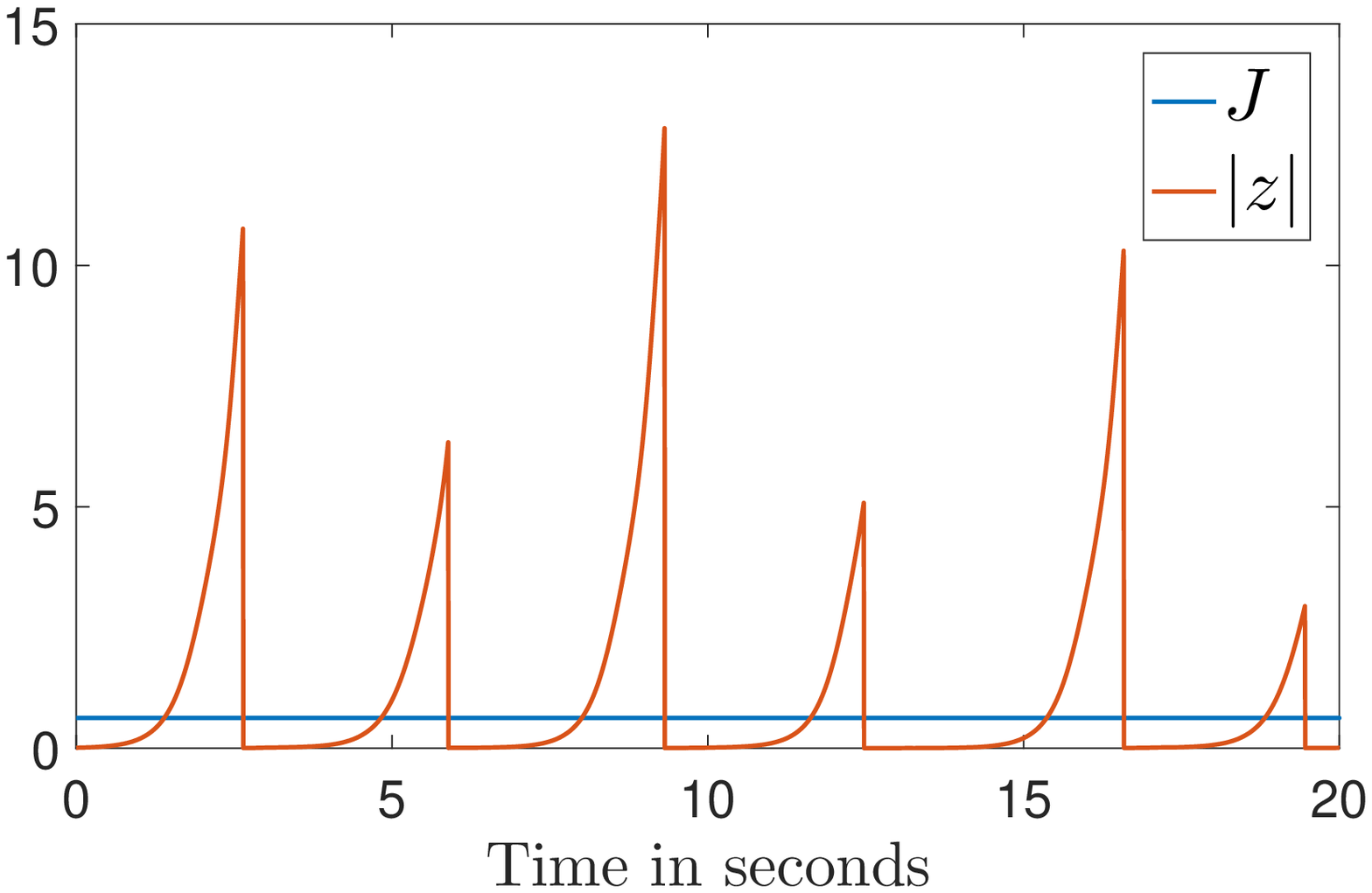} \\
    \includegraphics[width=60mm]{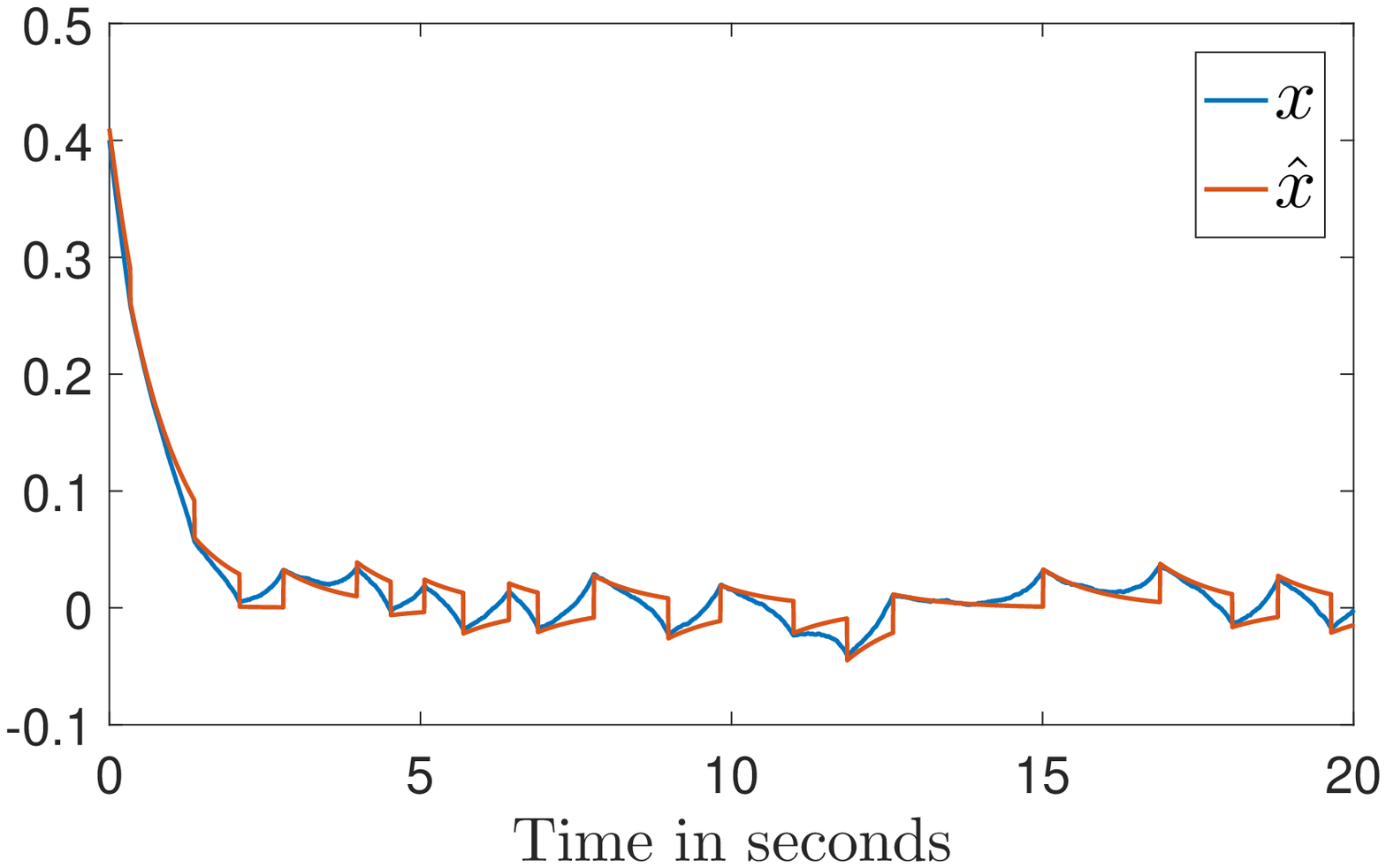} &
    \includegraphics[width=60mm]{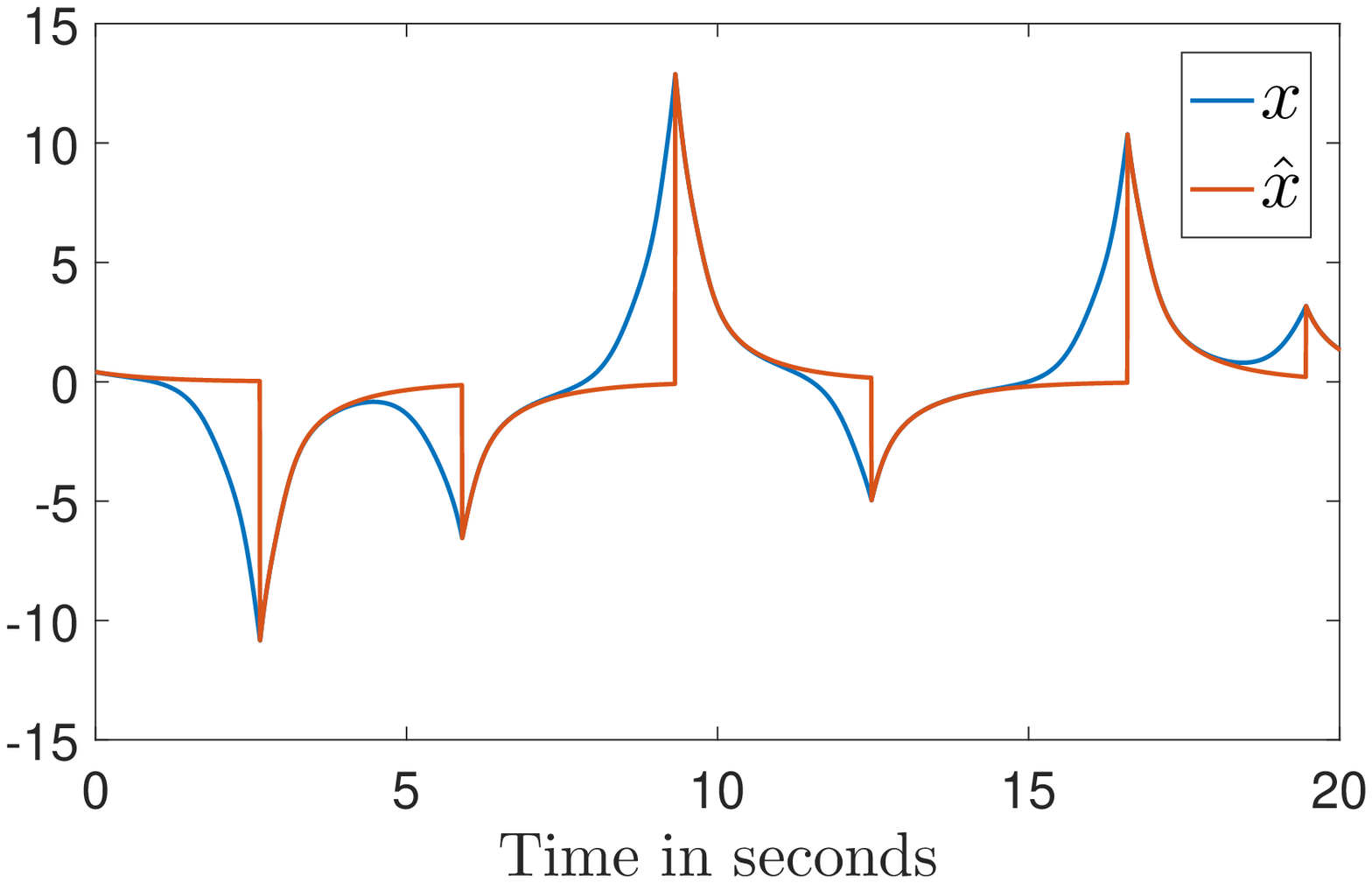} \\
\end{tabular}
\caption{Simulation results for stabilization of the plant~\eqref{nonlineasimulations}. We used the following parameters for the simulation:
sampling time $\delta=0.005$, simulation time $T=20$, $u(t)=-4\hat{x}(t)$, $\alpha=0.01$, packet size~\eqref{packetsizeforsimulation1294o312j}, and triggering threshhold $J=(e^{3 \gamma}-1)M/3+0.01$. 
}
\label{nonl_sim_res}
\end{figure*}
\begin{figure}[tbh]
  \centering
  \includegraphics[scale=0.22]{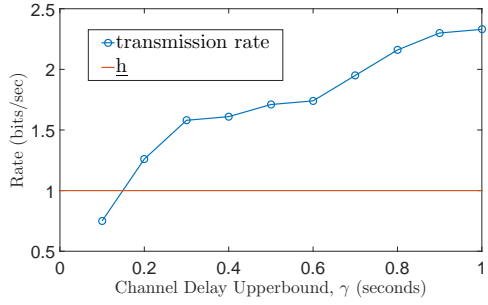} 
  \caption{
  Information transmission rate in simulations compared to the $\underbar{h}$~\eqref{entropylower-vec}.
  We used the following parameters for the simulation:
sampling time $\delta=0.01$ seconds, simulation time $T=100$ seconds, $u(t)=-2\hat{x}(t)$, $z(0)=0.01$, $M=0.05$, $\alpha=0.01$, packet size~\eqref{packetsizeforsimulation1294o312j}, and triggering threshhold $J=(e^{3 \gamma}-1)M/3+0.05$. 
  Note that the rate calculated from simulations can not start at $\gamma=0$ because the minimum channel delay upper bound is equal to two sampling time.}
  \label{ratenon3nonlinear}
\end{figure}
Here, we discuss some additional remarks about the nonlinear systems results presented in \secref{sec:nonlinear1}.
\begin{remark} {\rm 
Unlike the linear case, a closed form solution of~\eqref{nonlinearstateestii} is not known in general. Consequently, to simplify the encoding process, we use the  periodic event-triggering scheme~\eqref{triggerinperiod} and~\eqref{eq:etsnonlinear} (cf.~\cite{heemels2013periodic}), which is different from the  continues time event-triggering scheme~\eqref{eq:ets} where a triggering could occur at any time $t_s^k \ge 0$.
%. Here unlike the event-triggering scheme~\eqref{eq:ets} where a triggering could occurs at any time $t_s^k \ge 0$, a triggering could only happens at discrete time $t_s^k$ given in~\eqref{triggerinperiod}.
  }
  \oprocend
\end{remark}
\begin{remark} {\rm 
Although \asumref{ISS1} is restrictive, it is widely used in control of nonlinear systems under communication constraint~\cite{sharon2012input,de2004stabilizability,liberzon2005stabilization}. An exception is the work~\cite{de2004stabilizability} which eliminated this  assumption for systems without disturbances. An alternative ISS assumption which centers around state estimation $\hat{x}$ is proposed in~\cite{liberzon2005stabilization} where the evolution of state estimation $\hat{x}$ is described by an \textit{impulsive system}~\cite{hespanha2008lyapunov}. As in our event-triggering design the behavior of the state estimation $\hat{x}$ is described with an impulsive system~\eqref{nonlinearsystemestime} and~\eqref{estimNonlin}, the study of this alternative ISS assumption for our setup with a digital communication channel with bounded but unknown delay is an interesting research venue.
  }
  \oprocend
\end{remark}
 \begin{remark} {\rm 
The lower bound given on the packet size in~\eqref{nifekn2!!!!!2} might not be a natural number in general. This lower bound is used to properly bound the information transmission rate~\eqref{ratemo232oe4e} in~\eqref{inftranratenonlinear2}. In addition, the lower bound~\eqref{nifekn2!!!!!2} might be zero. When $g(t_s)=0$ there is no need to put any data payload in the packet and the plant can be stabilized using only timing information. However, in this case the sensor still needs to inform the controller about the occurrence of a triggering event. Consequently, when $g(t_s)=0$ is sufficient, the sensor can stabilize the system by transmitting a fixed symbol from a unitary alphabet to the controller (see~\cite{khojasteh2018stabilizing}). 
%one can assume when a triggering happens at time $t_s^k$ a fixed symbol from a unitary alphabet is transmitted to the controller (see~\cite{khojasteh2018stabilizing}). 
In practice, the packet size should be a natural number or zero, so if we do not want to use the fixed symbol from a unitary alphabet, as in~\eqref{packet_size}, the packet size
\begin{align}\label{packetsizeforsimulation1294o312j}
  g(t_s) = \max\left\{1,\left\lceil\log\left(\frac{\Upsilon(0)e^{L_x \gamma}}{J-\frac{L_wM}{L_x}\left(e^{L_x \gamma}-1\right)}\right) \right\rceil\right\},
  \end{align}
  is sufficient for stabilization.
  }
  \oprocend
\end{remark}
\begin{remark} {\rm As we used the trivial upper bound on the triggering rate $(\alpha+\gamma)^{-1}$ to deduce the bound~\eqref{inftranratenonlinear2}, this upper bound on $R_s$ might be too conservative in general.
  }
  \oprocend
\end{remark}
\begin{remark} {\rm %As mentioned in Definition~\ref{cd12emkmkd} 
When $\gamma=M=0$, 
%ISpS becomes equivalent to global asymptotic stability (GAS)~\cite{lin1996smooth}. In this case,
the data-rate theorem\cite{khojasteh2017time,princeton_paper,khojasteh2018event21122} states that the rate at which the controller receives information should be at least as large as the intrinsic entropy rate of the plant defined in~\cite{Topological}.   In our design, we can supply this information only using the implicit timing information in the triggering events. In fact, when $\alpha \rightarrow 0$ the periodic event-triggering control schemes~\eqref{triggerinperiod} and~\eqref{eq:etsnonlinear} become equivalent to the continuous time event-triggering policy~\eqref{eq:ets}. In this case, in a triggering time $t_s$ the controller can discover the exact value of $x(t_s)$ using equation $x(t_s)=\hat{x}(t_s)\pm J$ by receiving a single bit corresponding to the sign of $z(t_s)$. As there is no system disturbance, the controller then can track $x(t)$ using~\eqref{nonlinearsystemestime} after a single triggering time, and $R_s$~\eqref{ratemo232oe4e} will be arbitrarily small.  
  }
  \oprocend
\end{remark}
\subsection{Simulations for nonlinear systems}\label{sec:sim}
This section presents simulation results validating the proposed nonlinear scheme. While our analysis is for continuous-time plants, we perform the simulations in discrete time with a small sampling time~$\delta$. In this case, as discussed in \secref{sec:architect},  the minimum upper bound for the channel delay is equal to two sampling times.  We illustrate the execution of our design for the system
\begin{align}
\label{nonlineasimulations}
    \dot{x}=f(x(t),u(t),w(t))=2x(t)+\sin(x(t))+u(t)+w(t).
\end{align}
During inter-reception time, state estimation is defined according to~\eqref{nonlinearsystemestime}. Thus, using~\eqref{nonlinearstateestii}, for $t \in (t_c^k,t_c^{k+1})$ we deduce 
\begin{align}
    \label{zforsumnonli1123}
    \dot{z}(t)=2z(t)+\sin x - \sin \hat{x} +w(t).
\end{align}
Since $|\sin x - \sin \hat{x}| \le |x-\hat{x}|$, the dynamics~\eqref{nonlineasimulations} satisfies the Lipschitz property~\eqref{lishitssequ} with $L_x=3$, $L_w=1$ for all $|z(t)| \in \mathbb{R}_{\ge 0}$.

%Expriemnt 1: J, z, x , xhat
%$\gamma=0.99$
%$\alpha+\gamma=1$
%$x(0)=0.101$
%$xhat(0)=0.100$
%$M=0.1$, . To ensure~\eqref{conditiontocalculatee3444} we let 
%\begin{align}
%  g(t_s) = \max\left\{1,\left\lceil\log\left(\frac{\Upsilon(0)e^{3 \gamma}}{J-(e^{3 \gamma}-1)/30}\right) \right\rceil\right\},
%  \end{align}
%  where
%  \begin{align}
%    \Upsilon(0) =J e^{3(0.01+\gamma)}+ (e^{3(0.01+\gamma)}-1)/30.
%\end{align}
A set of two simulations are carried out for different values of $\gamma$ and $M$. Each column in \figref{nonl_sim_res}  presents one set of simulation. The first row shows the triggering threshold $J$  and the absolute value of the state estimation error $|z(t)|$. If the absolute value of this error is equal to $J$ during  the period $\alpha+\gamma$, the sensor transmits a packet at the end of this period, and the jumping strategy~\eqref{estimNonlin} adjusts $\hat{x}$ at the reception time to ensure the plant is ISpS. 

Note that the amount this error exceeds the triggering function depends on the random channel delay, upper bounded by $\gamma$. 
% which is upper bounded by $\gamma$. 
The second row of \figref{nonl_sim_res} presents the evolution of the state~\eqref{nonlineasimulations} and its estimation~\eqref{nonlinearsystemestime}. As expected, when $\gamma$ increases, while the plant remains ISpS the controller performance deteriorate significantly. 

As discussed in \secref{sec:expsetup}, according to  the data-rate theorem, to stabilize the plant, the information rate communicated over the channel in data payload and timing should be larger than the entropy rate of the plant~\cite{khojasteh2018stabilizing,khojasteh2017time}. Using~\cite{Topological} the entropy rate of the plant~\eqref{nonlineasimulations} at point $x^*$ is equal to $h(x^*)=\partial f/\partial x|_{x=x^*} =2+\cos(x^*(t))$. Thus, for any value of the state, the information accessible to the controller about the plant or the information rate communicated over the channel in data payload and timing, should be larger than 
\begin{align}
    \label{entropylower-vec}
    h(x)\ge \underline{h}=1. 
\end{align}
 \figref{ratenon3nonlinear} presents the simulation of information transmission rate versus the delay upper bound $\gamma$ in the communication channel to render~\eqref{nonlinearsystem1} ISpS. It can be seen that for small values of $\gamma$, the plant is ISpS with an information transmission rate smaller than the one prescribed by the data-rate theorem. Furthermore, as $\gamma$ increases, more information
has to be sent via data payload for stabilization since larger
delay corresponds to more uncertainties about the value of
the states at the controller and less timing information.
\end{document}